\documentclass[a4paper,11pt]{article}

\usepackage{graphicx}
\usepackage{amsmath,amsfonts,amssymb}
\usepackage{bbold}

\def\lta{~\mbox{\raisebox{-.6ex}{$\stackrel{<}{\sim}$}}~}
\def\gta{~\mbox{\raisebox{-.6ex}{$\stackrel{>}{\sim}$}}~}
\newcommand{\beq}{\begin{equation}}
\newcommand{\eeq}{\end{equation}}

\begin{document}

\begin{flushright}
UMN-TH-3219/13 \\
\end{flushright}

\vspace{1cm}

\begin{center}
{\LARGE \bf
Gauge-flation confronted with Planck  }\\[0.5cm]

{
{\large\bf Ryo Namba \,,\,
Emanuela Dimastrogiovanni \,,\,
Marco Peloso
}
}
\\[7mm]

 School of Physics and Astronomy,
University of Minnesota, Minneapolis, 55455, USA \\
\vspace{-0.3cm}

\vspace{1cm}

{\large\bf Abstract}

\end{center}
\begin{quote}

Gauge-flation is a recently proposed model in which inflation is driven solely by a non-Abelian gauge field thanks to a specific higher order derivative operator. The nature of the operator is such that it does not introduce ghosts. We compute the cosmological scalar and tensor perturbations for this model, improving over an existing computation. We then confront these results with the Planck data. The model is characterized by the quantity $\gamma \equiv \frac{g^2 Q^2}{H^2}$ (where $g$ is the gauge coupling constant, $Q$ the vector vev, and $H$ the Hubble rate). For $\gamma < 2$, the scalar perturbations show a strong tachyonic instability. In the stable region, the scalar power spectrum $n_s$ is too low at small $\gamma$, while the tensor-to-scalar ratio $r$ is too high at large $\gamma$. No value of $\gamma$ leads to acceptable values for $n_s$ and $r$, and so the model is ruled out by the CMB data. The same behavior with $\gamma$ was obtained in Chromo-natural inflation, a model in which inflation is driven by a pseudo-scalar coupled to a non-Abelian gauge field. When the pseudo-scalar can be integrated out, one recovers the model of Gauge-flation plus corrections. It was shown that this identification is very accurate at the background level, but differences emerged in the literature concerning  the perturbations of the two models. On the contrary, our results show that the analogy between the two models continues to be accurate also at the perturbative level.

\end{quote}


\section{Introduction}

\label{sec:intro}

The paradigm of inflation offers a successful theoretical framework for the physics of the Early Universe \cite{Linde:2005ht}. In most of its realizations, inflation is driven by scalar fields, a feature which can easily result in an isotropic expansion. Quantum fluctuations generated from scalar fields during or at the end of inflation can account both for the cosmological fluctuations in the Cosmic Microwave Background (CMB) Radiation and for the formation of Large Scale Structures. A large number of inflationary models with scalar fields provide results in excellent agreement with observations  \cite{Ade:2013uln}.  Nevertheless, models of inflation where vector fields play a major role have been receiving quite some attention, inspired both by theoretical and observationally-related motivations (see e.g. \cite{review1,review2,review3,review4,review5} for some recent reviews). 

In this paper, we study one of these models, namely  the recently proposed Gauge-flation model \cite{Maleknejad:2011jw}.  Gauge-flation is a non-Abelian gauge theory minimally coupled to gravity. At the background level, it is characterized by an effective scalar degree of freedom in the form of the vacuum expectation value (vev) of the spatial component of the gauge field. The action for the model reads \cite{Maleknejad:2011jw} 
\begin{equation}
\label{oneeq}
S=\int d^{4}x \sqrt{-g}\left[\frac{M_{p}^{2}}{2}R-\frac{1}{4}F_{\mu\nu}^{a}F^{a,\mu\nu}+\frac{\kappa}{96}\left(F_{\mu\nu}^{a}\tilde{F}^{a,\mu\nu}\right)^{2}\right],
\end{equation}
where $R$ is the Ricci scalar, $F_{\mu\nu}^{a} \equiv \partial_\mu A_\nu^a -  \partial_\nu A_\mu^a - g \epsilon^{abc} A_\mu^b A_\nu^c$ is the field-strength tensor of a SU(2) gauge field with coupling constant $g$,  and $\tilde{F}^{a,\mu\nu} \equiv \frac{\epsilon^{\mu\nu\alpha\beta}}{2 \sqrt{-g}} F_{\alpha\beta}^a$ is its dual ($\epsilon^{\mu\nu\alpha\beta}$ is totally anti-symmetric, and $\epsilon^{0123}=1$).

Thanks to its anti-symmetric structure, the last term in   (\ref{oneeq}) does not introduce more than two time derivatives in the equations of motion, and therefore, for positive $\kappa$, the model is ghost free.~\footnote{Ref. \cite{Elizalde:2012yk} studied at the background level  a model characterized by the last term in (\ref{oneeq}) plus a more generic kinetic structure ${\cal L } \supset {\cal F} \left[ - F^2 \right]$, where ${\cal F}$ is an arbitrary function.} Moreover, the gauge symmetry is not explicitly broken in this model, and there are no problems associated with the longitudinal vector polarizations, which instead destabilize several models of vector fields in cosmology  \cite{hcp}.

The model admits an isotropic background solution, with 
\begin{equation}
\label{gF}
\langle  A_{i}^{a} \rangle={\hat \phi } \left( t \right) \, \delta_{i}^{a}.
\end{equation}
Further studies of the model were carried out for example in \cite{Maleknejad:2011jr}, where the cosmic no-hair theorem is tested within Gauge-flation with a Bianchi I background, in \cite{Ghalee:2012gg}, where the background solutions of Gauge-flation were explored, and in \cite{Noorbala:2012fh}, where it is shown that the theory can be embedded in the gravi-leptogenesis scenario of \cite{Alexander:2004us}.  Recently, it was also shown \cite{Adshead:2012qe,SheikhJabbari:2012qf} that Gauge-flation shares some background trajectories  with Chromo-natural inflation \cite{Adshead:2012kp}, another recent model which assigns to non-Abelian gauge fields a crucial role in the dynamics of inflation and in the generation of primordial fluctuations (see  \cite{review4}  for a comprehensive review).~\footnote{See \cite{Martinec:2012bv} for an extension of Chromo-natural inflation.} 

 In Chromo-natural inflation an SU(2) gauge field is coupled to an axion $\varphi$ which plays the role of the inflaton:  
\begin{equation}
\label{cn}
S=\int d^{4}x\sqrt{-g}\left[\frac{M_{p}^{2}}{2}R-\frac{1}{4}F_{\mu\nu}^{a}F^{a,\mu\nu}-\frac{1}{2}\left(\partial\varphi\right)^{2}-V(\varphi)+\frac{\lambda}{4f}\varphi F_{\mu\nu}^{a}\tilde{F}^{a,\mu\nu}\right].
\end{equation}
It turns out that, if sufficiently close to the bottom of its potential, the $\varphi$ field can be integrated out, leading precisely to the action in Eq.~(\ref{oneeq}) plus corrections  \cite{SheikhJabbari:2012qf}. 

The main goal of  Chromo-natural model, and Gauge-flation as well,   is to solve an issue that affects 
   most scalar field models of inflation, i.e. the difficulty in ensuring a quasi-flat potential for the inflaton. The condition of quasi-flatness for the potential is important in order to allow enough e-foldings of expansion and to obtain a nearly scale invariant spectrum of perturbations.  This motivated the so-called \textsl{natural inflation} model \cite{Freese:1990rb,Adams:1992bn}, where the inflaton enjoys a (broken)  axionic shift symmetry, $\varphi\rightarrow\varphi+const.$, that protects its potential from  large quantum corrections.   The simplest implementation of this idea, namely a single field $\varphi$ evolving in a potential $V \propto \cos \frac{\varphi}{f}$,   agrees with observations  only  if the axion decay constant $f$ is of the order of or greater than the Planck mass \cite{Savage:2006tr}. It has been debated in the literature whether a trans-Planckian breaking may be compatible with gravity, in the case in which the symmetry is global \cite{f-Mp-global}.  The shift symmetry may emerge from a gauge symmetry,  as typically in string theory. However, also in this case a trans-Planckian $f$ is  regarded as  problematic, since all known  controlled string theory constructions are characterized by   $f<M_{p}$   \cite{Banks:2003sx,Svrcek:2006yi}.

A scale  $f < M_p$ can be compatible with inflation through a number of mechanisms, several of which  also lead   to an interesting phenomenology  (see     \cite{review5} for a recent  review).   One can, for instance, consider more than one axion \cite{sol1,sol2}, require nontrivial compactifications in string theory  \cite{sol3}, couple the axion to a $4$-form \cite{sol4}, modify the axion kinetic term \cite{sol5}, or slow-down the axion through particle production \cite{sol6a,sol6b}. In particular, in the mechanism of \cite{sol6a} the dissipation occurs through the production of a U(1) field coupled to the inflaton $\varphi$ through the interaction $\frac{\varphi}{f} F {\tilde F}$ between the inflaton and an Abelian gauge field.  Ref.  \cite{Adshead:2012kp}  showed that this coupling can also affect the background evolution (prior to any particle production consideration) if the U(1) field is replaced by a SU(2) field with a nonvanishing vacuum expectation value (vev). Due to the interaction with the vev of the vector multiplet, the inflaton can be in slow roll even if its potential would otherwise (i.e. in absence of this interaction) be too steep to give inflation. As shown in  \cite{Adshead:2012qe,SheikhJabbari:2012qf}, Gauge-flation appears as an analogous (``dual'') version of this mechanism. The Gauge-flation formulation is particularly suggestive since it is  characterized by the vector field only.   To our knowledge, this is the first and only existing  stable model in which inflation is driven by a vector field alone (previously introduced models with dynamical vector fields during inflation either  have dynamically relevant scalar fields \cite{review4}, or  ghosts \cite{hcp}).

The perturbation analysis for Chromo-natural inflation was first carried out in \cite{Dimastrogiovanni:2012st} in the  limit of heavy vector field (see below), in which the gauge field can be integrated out, leading to an effective single-scalar field $P[(\partial\varphi)^{2},\varphi]$  model of inflation with a non-canonical kinetic term. A full analysis at linear order in perturbation theory was later presented in \cite{Dimastrogiovanni:2012ew}, where the model was found to be highly  unstable  in the sub-horizon regime for: 
\begin{equation}
\frac{g^2 \, Q^2}{H^2} < 2 \;\;\;,\;\;\; {\rm instability \; in \;   Chromo\mbox{-}natural \; inflation} \;\;. 
\label{cn-instab}
\end{equation}
In this relation $Q$ is  the SU(2) vev, related to (\ref{gF}) by $Q = {\hat \phi } / a$ (where $a$ is the scale factor of the universe), while $H$ is the Hubble rate.  In Chromo-natural inflation, the quantity $m_g \equiv \sqrt{2} \, g \, Q$ coincides with the mass of the vector field fluctuations in the $m_g \gg H$ regime  \cite{Dimastrogiovanni:2012st},
which is the regime where the computation of  \cite{Dimastrogiovanni:2012st} applies.

Ref.   \cite{Dimastrogiovanni:2012ew}, also noted  that the model can lead to a large production of the  vacuum gravity wave mode, in excess of the standard Lyth bound \cite{Lyth:1996im}  $r > 16 \epsilon$ (see eqs. (\ref{epsilon}) and (\ref{def-r}) for the definition of $\epsilon$ and $r$, respectively), which does not apply in this context, since it holds for a free inflaton and unsourced tensor modes. The stability result  (\ref{cn-instab})   was later obtained independently by  \cite{Adshead:2013qp}, that  also presented a complete study of the gravitational waves produced in the model (correcting an error in the original version of  \cite{Dimastrogiovanni:2012ew}), showing that they are chiral. Ref.  \cite{Dimastrogiovanni:2012ew} also presented the scalar and tensor power spectrum of the Chromo-natural inflation for some illustrative choices of the parameters. None of the examples presented in \cite{Dimastrogiovanni:2012ew} leads to an acceptable phenomenology. In particular, these example showed that the spectrum of the scalar perturbations is 
too red at small $m_g/H$, while the tensor modes are too high at large $m_g/H$. Based on this, ref.  \cite{Dimastrogiovanni:2012ew} argued that the simultaneous requirements of sufficiently flat scalar spectrum and of sufficiently small tensor mode could pose significant bounds on the model, and potentially rule it out as a model of sub-Planckian $f$. This was later confirmed by \cite{Adshead:2013nka} through an exhaustive parameter scan in the model.

A study of the  perturbations   of Gauge-flation can instead be found in   \cite{Maleknejad:2011jw}.  Ref.  \cite{Maleknejad:2011jw} concluded that all parameter choices lead to a stable solution (contrary to what happens in Chromo-natural inflation, see eq. (\ref{cn-instab})). Moreover, the scalar spectrum obtained in   \cite{Maleknejad:2011jw} is significantly bluer than the one found  for Chromo-natural inflation;  compare for instance Figure 8 of the last work in   \cite{Maleknejad:2011jw}  with  Figure 12 of  \cite{Adshead:2013nka}. This different behavior is somewhat puzzling, given the analogy that the two models present at the background level. Specifically, ref.  \cite{SheikhJabbari:2012qf} shows that, integrating out the axion  $\varphi$
from the action (\ref{cn}) of Chromo-natural inflation,   one obtains the action (\ref{oneeq}) of Gauge-flation, plus corrections.~\footnote{To be precise, integrating out $\varphi$ is  feasible only when  $\varphi$ is close to the minimum of its potential; so, when the two models are compared to each other, one is actually identifying Gauge-flation with a specific limit of Chromo-natural inflation. For brevity of exposition, we refer to the two models as being ``analogous'', or ``dual'', but we stress that Chromo-natural inflation has actually  a larger parameter space than Gauge-flation \cite{Adshead:2012qe,SheikhJabbari:2012qf}.}
 The procedure is conceptually simple, but it is not straightforward to compute the corrections explicitly. The integration of $\varphi$ is done in \cite{SheikhJabbari:2012qf} at the level of the one loop effective action;  slow roll approximations are used, and in computing the functional determinant the laplacian has been replaced by its flat space value, under the assumption that the main contribution comes from sub-horizon momenta   \cite{SheikhJabbari:2012qf}.  One may wonder about the impact  of the corrections that would emerge by going beyond these approximations.  For instance, the  departure of the perturbations from scale invariance is controlled by the full evolution (and not only the sub-horizon regime) and by the slow roll parameters. Barring mistakes,  the difference  emerged in the literature   between the perturbations in the two models  is a clear indication  that a non-negligible  difference between the two models must appear when $\varphi$ is accurately integrated out.

It is interesting to understand how precisely this difference emerges, as this may provide some general indication on 
how accurate one needs to be in integrating out  fields from a model of inflation, and still obtain an accurate effective description. This was the main motivation for the present study.  As a byproduct, we also deemed  useful to 
 update the phenomenological limits on Gauge-flation in the light of the new Planck results \cite{Ade:2013zuv}, 
particularly if the analogy with Chromo-natural inflation turns out to be accurate also at the perturbative level. 
 To understand   this,   we computed the cosmological perturbations of  Gauge-flation, employing  an analogous procedure to the one  that we used to study the perturbations of Chromo-natural inflation \cite{Dimastrogiovanni:2012ew}.  Disregarding the vector perturbations (as we realized that considering them would not impact our conclusions) the model has two physical scalar perturbations, two physical left-handed tensor perturbations, and two physical right-handed perturbations. These three groups are decoupled from each other at the linearized level, and each of them is described by two coupled second order differential equations.  The equations (particularly, those in the scalar sector), are extremely involved and we could not solve them analytically. 
 However, they can be  integrated numerically without any particular difficulty.  After imposing a given duration of inflation (we study both the cases in which there are $N=50$ or $N=60$ e-folds between the moment when the largest CMB 
modes exit the horizon and the end of inflation), and that the scalar perturbations have the observed amplitude, Gauge-flation is characterized by a single free  parameter, that can be chosen to be the same combination $\gamma \equiv g^2 \, Q^2 / H^2$ that also plays a relevant role for the perturbations of Chromo-natural inflation. We obtained phenomenological results for Gauge-flation as a function of this parameter.  Our results differ from those of \cite{Maleknejad:2011jw}, and agree with those emerged from the several analyses of Chromo-natural inflation. In particular, we find that the stability and departure from scale invariance of the scalar modes, and the level of the tensor modes of Gauge-flation scale with  $\gamma$ analogously to what found in  Chromo-natural inflation. We find that  also  Gauge-flation is ruled out by the CMB data.

The paper is organized as follows. In Section \ref{sec:model} we introduce the model and review the background analysis, mostly summarizing the analogous computation of   \cite{Maleknejad:2011jw}.  In Section \ref{sec:linpert} we present our formalism for the perturbations. Specifically, we discuss at the formal level  how we fix the gauge freedom, integrate out the  non-dynamical modes, and quantize the early time action, to obtain the initial condition for the perturbations. The explicit computations for the model are presented in Section \ref{sec:tensor} for the tensor perturbations, and in Section \ref{sec:scalar} for the scalar modes. In  Section \ref{sec:pheno} we study the phenomenological implications of these results. In Section \ref{sec:comparison} we compare our computation with that of  \cite{Maleknejad:2011jw}. Finally, in the concluding  Section \ref{sec:conclusions} we summarize our findings, and compare them with those in Chromo-natural inflation.


\section{The model, the background solution, and the slow roll approximation}

\label{sec:model}

The ``matter'' Lagrangian of Gauge-flation, see eq.~(\ref{oneeq}), consists of a standard Yang-Mills term and of a gauge-invariant contribution of the form $(F\tilde{F})^{2}$ ($\tilde{F}$ being the dual field-strength tensor). 
 Due to the SU(2) gauge symmetry, the theory admits a rotationally invariant vev for the SU(2) multiplet, 
  specified by eq.~({\ref{gF}}), which is compatible with an isotropic  background.  We choose an FRW background metric \footnote{Notation: in this work, dot denotes derivative with respect to physical time $t$, while prime denotes derivative with respect to conformal time $\tau$, related to the physical time by $d t = a \, d \tau$. Greek indices span all coordinates, while latin indices span the spatial coordinates.}
\begin{equation}
\label{s2one}
ds^{2}=-dt^{2}+a^{2}(t)\delta_{ij}dx^{i}dx^{j}.
\end{equation}

The energy-momentum tensor of the system, 
\begin{equation}
T_{\mu \nu} =  F^{a}_{\mu \alpha} F^a_{\nu \beta} g^{\alpha \beta} - g_{\mu \nu} \left[  \frac{1}{4} F^2 + \frac{\kappa}{96} \left( F {\tilde F} \right)^2 \right] \;\;, 
\label{Tdd}
\end{equation}
once evaluated on the background, assumes the standard form  
\begin{equation}
\label{s2two}
T_{\;\; \nu}^{\mu}=diag\left(-\rho,P,P,P\right) \;\;. 
\end{equation}
In this expression, the energy density and the pressure can be written as the sum of two contributions, the first one arising from the Yang-Mills terms and the second one from the $(F\tilde{F})^{2}$ interaction
\begin{equation}
\label{s2three}
\rho=\rho_{YM}+\rho_{\kappa},\quad\quad\quad P=\frac{1}{3}\rho_{YM}-\rho_{\kappa} \;\;, 
\end{equation}
where
\begin{equation}
\label{s2four}
\rho_{YM}\equiv\frac{3}{2}\left(\frac{\dot{{\hat \phi } }^{2}}{a^{2}}+g^{2}\frac{{\hat \phi } ^{4}}{a^{4}}\right),\quad\quad\quad \rho_{\kappa}\equiv\frac{3}{2}\kappa g^{2}\frac{\dot{{\hat \phi } }^{2}{\hat \phi } ^{4}}{a^{6}} \;\;. 
\end{equation}
 We can immediately see that the standard  Yang-Mills contribution leads to the equation of state of radiation, while the second contribution to an effective equation of state of a cosmological constant.  The $(F\tilde{F})^{2}$ term  thus turns out to be a convenient choice for building a model of inflation. Despite being higher order, this term has only two time derivatives, due to its anti-symmetric structure, and the theory can be stable for some choice of parameters (as our study below shows). However, when this term dominates over the Yang-Mills one, it is natural to wonder whether higher order operators (which may arise from loops) can be safely suppressed.   The authors of \cite{Maleknejad:2011jr} claim that this is the case  within the parameter space of the theory. 

The background evolution of Gauge-flation has been exhaustively studied in the literature, starting from the original proposal  \cite{Maleknejad:2011jw}. We  summarize it here for completeness,  and to discuss the  freedom in the choice of parameters and initial conditions in the theory.

We can immediately verify that slow roll inflation requires (the standard FRW equations for $H$ and $\dot{H}$ apply)
\begin{equation}
 \epsilon \equiv - \frac{\dot{H}}{H^2} = \frac{2 \rho_{YM}}{\rho_{YM} + \rho_\kappa}  \ll 1  \;\; \Leftrightarrow \;\; \rho_{\kappa}\gg \rho_{YM}  \;\;,
\label{epsilon}
 \end{equation}
confirming that the energy associated to the higher order term in (\ref{oneeq}) must dominate over the one from the conventional Yang-Mills term. Besides the slow roll parameter  $\epsilon$, it is useful to introduce the two additional slow-roll parameters:
 \begin{equation}
 \label{s2seven}
  \eta\equiv -\frac{\ddot{H}}{2H\dot{H}}=\epsilon-\frac{\dot{\epsilon}}{2\epsilon H} \;\;\;,\;\;\;
 \delta\equiv -\frac{\dot{Q}}{HQ} \;\;.
 \end{equation}
 The parameter $\eta$ is related to the time variation of $\epsilon$, so that   $\eta \ll 1$ imposes that $\epsilon$ varies very slowly during inflation;  the other slow-roll parameter is defined from the time variation of the physical field
 \begin{equation}
\label{s2eight}
 Q \equiv \frac{{\hat \phi } }{a}  \;\;,
 \end{equation}
so   $  \delta \ll 1$ ensures that the gauge inflaton $Q$ sustains inflation long enough. The condition $\epsilon\ll 1$ in turn translates into $\kappa g^{2}Q^{4}\gg 1$ and $\kappa H^{2}Q^{2}\left(1-\delta\right)^{2}\simeq\kappa H^{2}Q^{2}\gg 1$. The slow-roll parameters satisfy the relations
\begin{equation}
\label{sl2}
\epsilon\simeq\frac{Q^{2}}{M_{p}^{2}}\left(1+\gamma\right),\quad\quad\eta\simeq\frac{\epsilon}{1+\gamma}\simeq\frac{Q^{2}}{M_{p}^{2}},\quad\quad\delta\simeq\frac{\gamma}{6\left(1+\gamma\right)}\epsilon^{2},
\end{equation}
where $\gamma $ is defined in (\ref{def-gamma}). The relations (\ref{sl2}) have been derived in  \cite{Maleknejad:2011jw} and for brevity we do not repeat the derivation here. We note that,  for the slow-roll conditions to be satisfied,   $Q\ll M_{p}$ is needed, so that, according to standard terminology, this is a   ``small-field'' inflationary model. However, as we  show below, contrary to  what happens in standard small-field models, Gauge-flation can lead to too large an amount of gravitational waves for some choice of parameters.

Following the notation of  \cite{Maleknejad:2011jw},  in eq. (\ref{sl2}) we have introduced the quantity 
\begin{equation}
\gamma \equiv \frac{g^{2}Q^{2}}{H^{2}} \;\;, 
\label{def-gamma}
\end{equation}
which we see corresponds to the same combination (\ref{cn-instab}) bounded by the stability of the scalar modes in  Chromo-natural inflation. One of the goals of the current work is to study whether a bound analogous to  (\ref{cn-instab}) is also present in Gauge-flation. 

By combining the definition (\ref{def-gamma})  with  eqs. (\ref{epsilon}) and (\ref{s2seven}) we can obtain the exact relation
\begin{equation}
\kappa   = \frac{1}{H^2 \gamma Q^2} \;  \frac{\left( 1 - \delta \right)^2 + \gamma}{\left( 1 - \delta \right)^2} \, \frac{2-\epsilon}{\epsilon} \;\;,
\label{relation-kappa}
\end{equation}
which  will be used in the following Sections  to eliminate the parameter  $\kappa$ from the explicit form of the equations of the perturbations that we solve. Another useful exact  relation that we will use in the study of the perturbations is
\begin{equation}
 M_p = Q \sqrt{\frac{\left( 1 - \delta \right)^2 + \gamma}{\epsilon}} \;\;.
\label{relation-Mp}
 \end{equation}
To prove this relation, we rewrite (\ref{epsilon}) as $\epsilon = \frac{2 \rho_{YM}}{3 H^2 M_p^2}$,  we express $\rho_{YM}$ through eqs. (\ref{s2four}), (\ref{s2eight}), and, finally, we trade $\dot{Q}$ for $\delta$ - using (\ref{s2seven}) - and $g^2$ for $\gamma$ - using (\ref{def-gamma}).

Differentiating (\ref{def-gamma}), and using the definition of the slow roll parameters, we obtain that
\begin{equation}
\dot{\gamma}= 2 \gamma \left(\epsilon-\delta \right) H \;\;,
\end{equation}
so that $\gamma$ is a slowly rolling quantity. More in general, from the slow roll conditions, one can show  \cite{Maleknejad:2011jw}  that  the quantities  $\gamma H^2 \propto Q^2 \propto \epsilon / \left( 1 + \gamma \right)$ are constant at leading order in slow roll.  Therefore,
\begin{equation}
\label{s2fourteen}
\frac{\epsilon}{\epsilon_{\rm in}}\simeq\frac{1+\gamma}{1+\gamma_{\rm in}}
\;\;,\;\;
\frac{H^{2}}{H_{\rm in}^{2}}\simeq \frac{\gamma_{\rm in}}{\gamma} \;\;,
\end{equation}
where the subscript ``in'' indicates some initial time, which we take  to correspond to  $N$ e-folds before the end of inflation. 

The number of e-folds can be related to the initial values of $\gamma$ and $\epsilon$ by \cite{Maleknejad:2011jw}  
\begin{equation}
\label{nN}
N\simeq \frac{1+\gamma_{\rm in}}{2\epsilon_{\rm in}}\ln\left[\frac{1+\gamma_{\rm in}}{\gamma_{\rm in}}\right].
\end{equation}
up to subleading terms in slow roll.

We introduce the  dimensionless quantities
\begin{equation}
{\tilde Q} \equiv \frac{Q}{M_p} \;\;,\;\; {\tilde t} \equiv \frac{t}{\sqrt{\kappa} M_p} \;\;,\;\;
{\tilde g} \equiv \sqrt{\kappa} M_p^2 g \;\;.
\label{tilde-combo}
\end{equation}
and we denote ${\tilde \partial} \equiv \frac{\partial }{ \partial {\tilde t} } $. In terms of these quantities, the only nontrivial SU(2) background equation and the background ``ii-Einstein'' equation  read, respectively
\begin{eqnarray}
& & 
 {\tilde \partial}^2 {\tilde Q} + 2 \frac{ {\tilde \partial} a}{a}  {\tilde \partial} {\tilde Q} +  \frac{ {\tilde \partial}^2 a}{a} {\tilde Q}  + \frac{  2 {\tilde g}^2 }{  1 + {\tilde g}^2 {\tilde Q}^4  }
\left[ 1 + \left( {\tilde \partial} {\tilde Q} + \frac{ {\tilde \partial} a}{a}  {\tilde Q} \right)^2 \right] {\tilde Q}^3 
+ \frac{ {\tilde \partial} a }{a} \frac{  1 - 3 {\tilde g}^2 {\tilde Q}^4  }{  1 + {\tilde g}^2 {\tilde Q}^4   } 
 \left(  {\tilde \partial} {\tilde Q} + \frac{ {\tilde \partial} a}{a}  {\tilde Q} \right)  = 0 \;\; , \nonumber\\ 
& & 
\frac{{\tilde \partial}^2 a}{a}  - \left( \frac{{\tilde \partial} a}{a}  \right)^2 = - \left[  \left(  {\tilde \partial} {\tilde Q} + \frac{ {\tilde \partial} a}{a}  {\tilde Q} \right)^2 + {\tilde g}^2 {\tilde Q}^4 \right] \;\; ,
\label{eom-Q-ii}
\end{eqnarray}
while the background  "00-Einstein'' equation reads 
\begin{equation}
 \left( \frac{{\tilde \partial} a}{a}  \right)^2 = \frac{1}{2} \left[ \left( {\tilde \partial} {\tilde Q} + \frac{ {\tilde \partial} a}{a}  {\tilde Q} \right)^2  \left( 1 + {\tilde g}^2 {\tilde Q}^4 \right)   + {\tilde g}^2 {\tilde Q}^4 \right] \;\;.
\label{eom-00}
 \end{equation}
These three equations are related to each other by a nontrivial Bianchi identity. A set of independent equations is provided by either one of (\ref{eom-Q-ii}) and by (\ref{eom-00}). Equivalently, one can solve the two equations 
(\ref{eom-Q-ii}), but then eq. (\ref{eom-00}) has to be imposed as an initial condition.  Whatever the choice, the background evolution  is completely determined by the initial  values  ${\tilde Q}_{\rm in}$ and  ${\tilde \partial } {\tilde Q}_{\rm in}$ (for a flat universe, the normalization of the scale factor is arbitrary, and we set $a_{\rm in} = 1 $). 
 
We now show that all quantities needed to specify a background evolution can be given in terms of $N$ and of $\gamma_{\rm in}$.   Eliminating $\epsilon$ through (\ref{sl2}),  eq. (\ref{nN}) becomes a relation in terms of $N, \gamma_{\rm in}$
 and  ${\tilde Q}_{\rm in}$, that we solve numerically to obtain  ${\tilde Q}_{\rm in}$ in terms of the other two quantities:
 \begin{equation} 
{\tilde Q}_{\rm in}  \simeq  \left[ \frac{1}{2 N}\ln\left(\frac{1+\gamma_{\rm in}}{\gamma_{\rm in}}\right) \right]^{1/2}  \;\;. 
\label{getQin}
\end{equation}
Next, consider  eq. (\ref{def-gamma}), and combine  eqs. (\ref{s2seven}) and (\ref{sl2}). These relations, written  in terms of dimensionless quantities, and evaluated at the initial time, give 
\begin{eqnarray}
&& {\tilde g} = \frac{ \gamma_{\rm in}^{1/2} {\tilde \partial } a_{\rm in} }{ {\tilde Q}_{\rm in} } \;\;, \nonumber\\ 
&& {\tilde \partial} {\tilde Q}_{\rm in} = - \delta_{\rm in} {\tilde Q}_{\rm in} {\tilde \partial a}_{\rm in}  \simeq - \frac{\gamma_{\rm in} \left( 1 + \gamma_{\rm in} \right) {\tilde Q}_{\rm in}^5}{6} \, 
 {\tilde \partial a}_{\rm in}  \;\; , 
\label{getgQp}
\end{eqnarray}
These two equations, plus  eq. (\ref{eom-00}) are three relations in terms of three unknown quantities ${\tilde g} ,\,  
 {\tilde \partial } a_{\rm in} ,\, $ and $  {\tilde \partial} {\tilde Q}_{\rm in} $. In this way we obtain all the quantities needed   
 for the background evolution.

To summarize: Gauge-flation is characterized by the two parameters $\kappa$ and $g$. A given inflationary evolution is specified by  the number of e-folds $N$,  the values of the inflaton ($Q$) and its time derivative at some initial time. One of the parameters ($\kappa$, in the current case) can be rescaled  out of the background evolution by reabsorbing it in the units of time (eq. (\ref{tilde-combo}), in the current case). We will determine this parameter by imposing that the scalar perturbations have the measured amplitude (this is completely analogous to what happens to the parameter $m$ in massive chaotic inflation, $V = \frac{1}{2} m^2 \varphi^2$). For any choice of the remaining parameter ${\tilde g}$, the initial value of the inflaton is one-to-one related to the number of e-folds of inflation. The initial  derivative of the inflaton is instead obtained by imposing that the evolution is in the slow roll inflationary attractor (in our case, the inflationary attractor is characterized by the last of (\ref{getgQp})).

\begin{figure}
\centering
\includegraphics[width=0.45\textwidth]{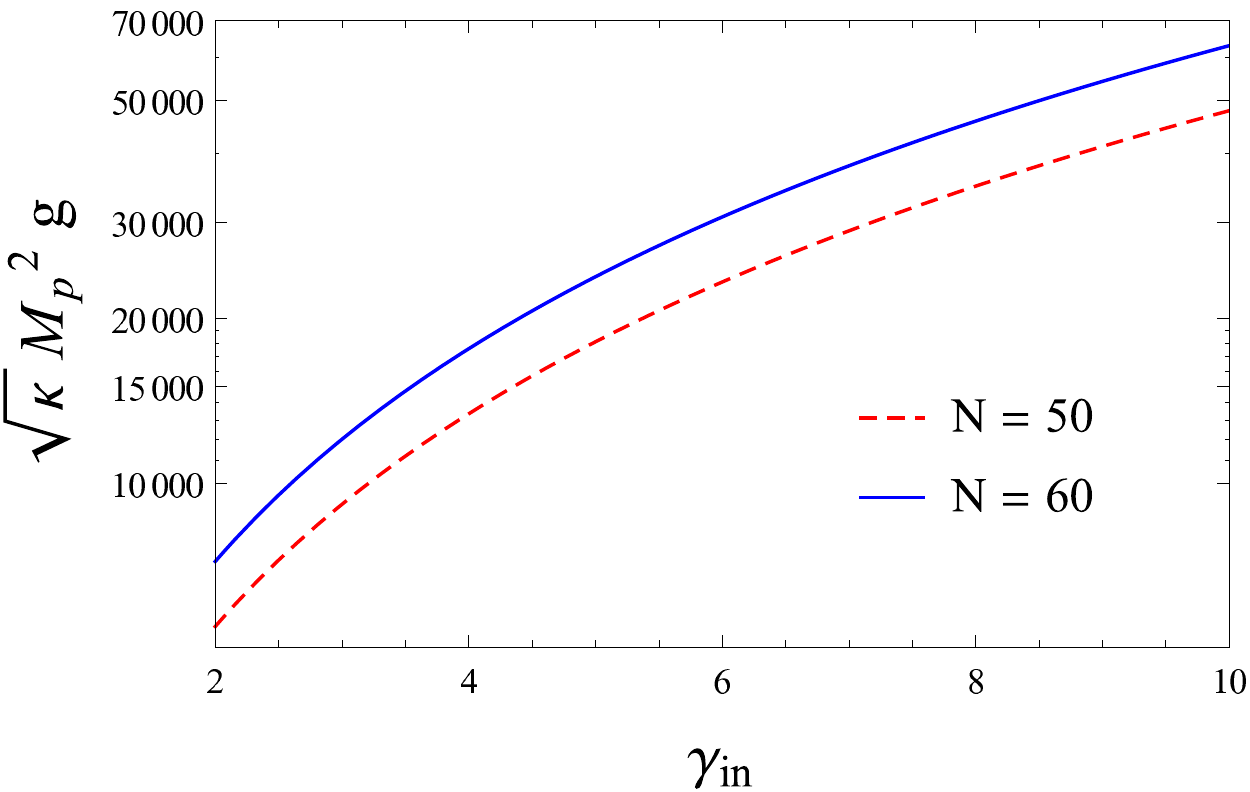}
\includegraphics[width=0.45\textwidth]{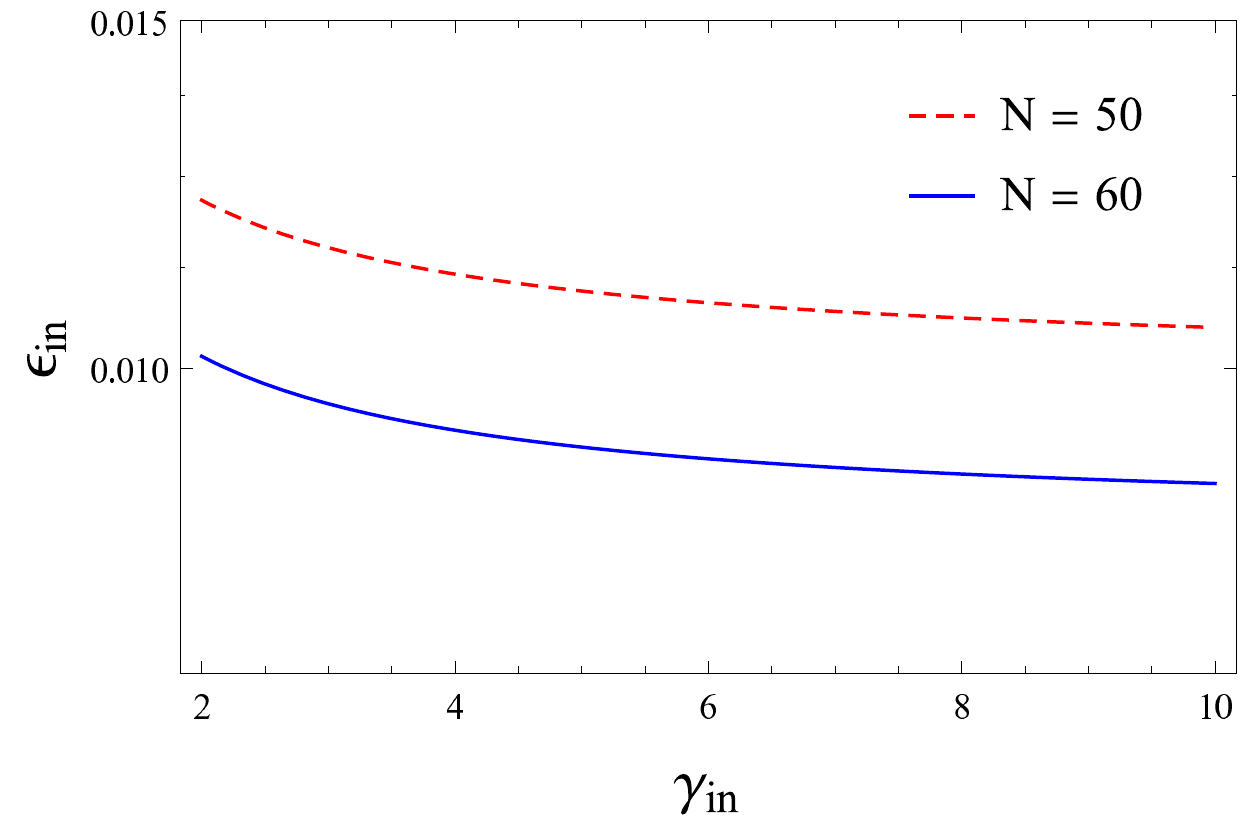}
\caption{Left panel: Relation between the dimensionless parameter ${\tilde g} \equiv \sqrt{\kappa} M_p^2 g $, and the initial value of $\gamma $, for $N=50$ and $N=60$ e-folds of inflation. The quantities $N$ and $\gamma_{\rm in}$
completely characterize the model and the background evolution. Right panel: relation between the initial value of the slow roll parameter $\epsilon$ and of $\gamma$.   \label{g-eps}}
\end{figure}

In our analysis, we ``traded'' the remaining parameter ${\tilde g}$  for $\gamma_{\rm in}$ by imposing the first of  (\ref{getgQp}). The quantity  $\gamma$  is constant at leading order in slow roll , so its initial value $\gamma_{\rm in}$ is a good quantity to characterize a given evolution. The choice of presenting our results in terms of  $\gamma_{\rm in}$ rather than ${\tilde g}$  is dictated by the fact that $\gamma$ played an important role in the related model of Chromo-natural inflation (see eq.  (\ref{cn-instab})).    However, we could have equivalently used ${\tilde g}$, and the left panel of Figure \ref{g-eps}  shows   how these two quantities are related for $N=50$ and $N=60$. In the right panel of the Figure we show instead how the initial value of the slow roll parameter $\epsilon$ is related to $\gamma_{\rm in}$.
    
\begin{figure}
\centering
\includegraphics[width=0.45\textwidth]{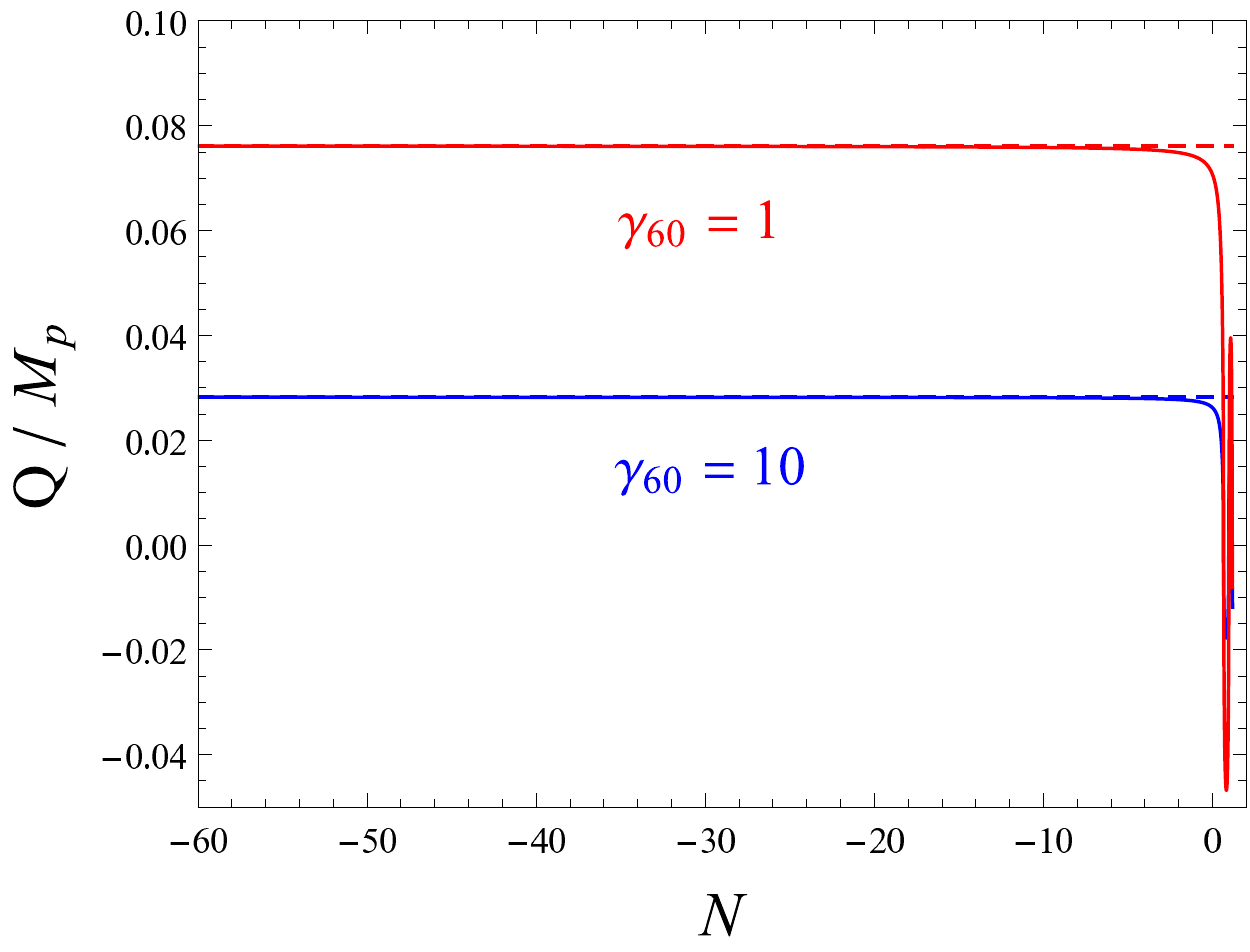}
\includegraphics[width=0.45\textwidth]{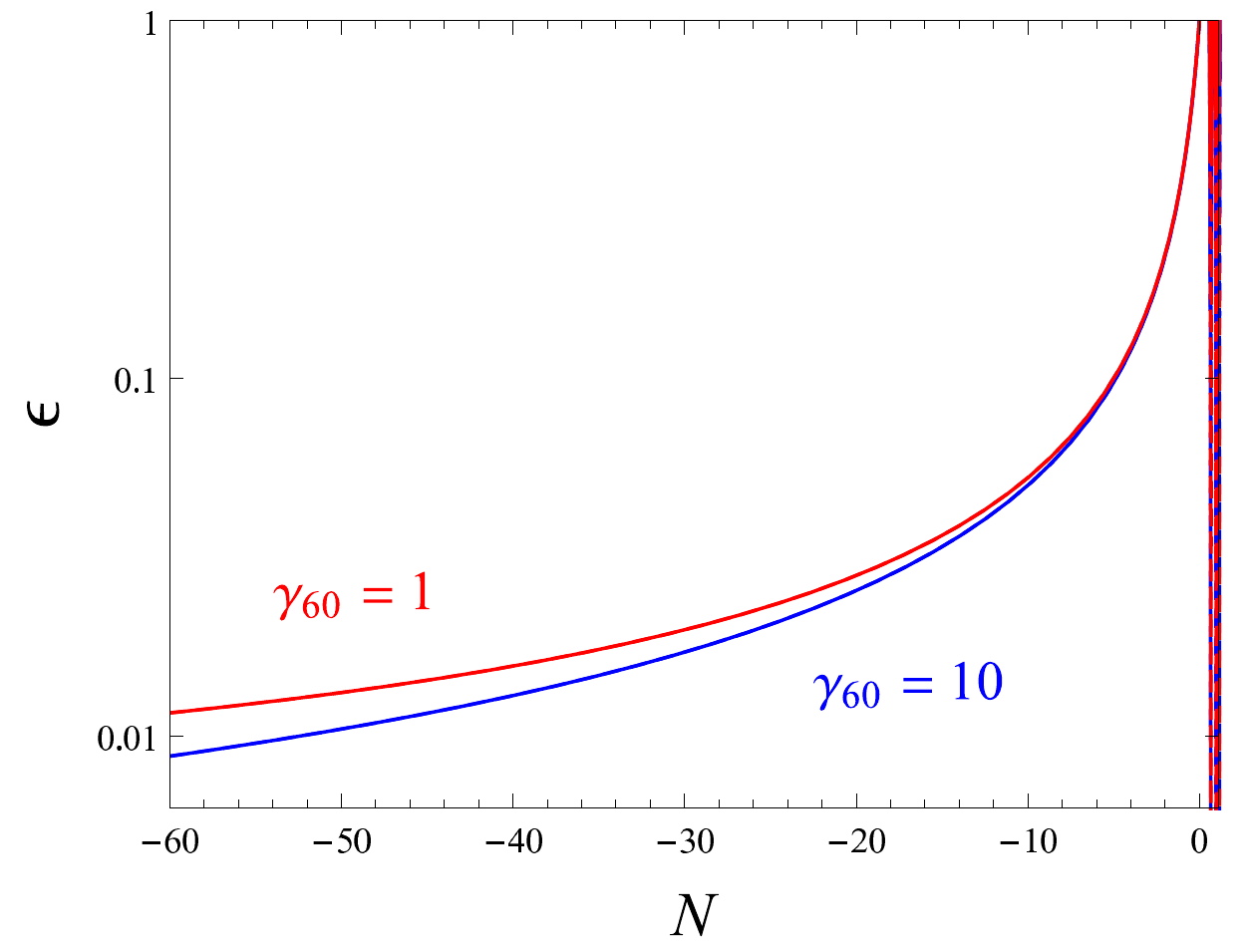}
\caption{ Inflationary evolution of $Q$ (left panel) and of $\epsilon$ (right panel) as a function of the number of e-folds, for  two different values  of $\gamma_{60}$ (the value $\gamma_{\rm in}$ at $60$ e-folds before the end of inflation).   Solid lines in the Figure give the numerical  exact evolution, while dashed lines the analytic slow-roll approximation. The solid and dashed  lines are almost  superimposed to each other,  signaling the great accuracy of the slow roll approximation.             
}
\label{fig:Q-eps-bck}
\end{figure}

 In Figure~\ref{fig:Q-eps-bck}, we show both the inflationary evolution of $Q(t)$ (left panel)  and of the slow roll  parameter $\epsilon$ (right panel), for $60$ e-folds of inflation and for two values of the parameter $\gamma$. 
 Solid lines in the Figure give the numerical  exact evolution, while dashed lines the analytic slow-roll approximation. 
The solid and dashed  lines are almost  superimposed to each other,  signaling the great accuracy of the slow roll approximation.

\section{Linear perturbations}

\label{sec:linpert}

In this Section, we study at a formal level the most general perturbations of the background solution presented in Section \ref{sec:model}. In Subsection \ref{subsesec:gauge} we split the perturbations  in three groups that are decoupled from each other at the linearized level (and that can be therefore studied independently), and we fix the gauge freedoms associated with general coordinate transformations and with the SU(2) group. In Subsection \ref{subsec:formal-deco} we present our formalism to solve the linearized equations for these perturbations.

\subsection{General decomposition and gauge choice} 

\label{subsesec:gauge}

This discussion closely follows  Subsection III A of \cite{Dimastrogiovanni:2012ew}, where the reader is referred to for details.  There are 22 perturbations in the model, $12$ of which in the SU(2) vector field, and $10$ in the metric. 
We decompose:
\begin{eqnarray}
A_0^a & = & a \left( Y_a + \partial_a Y \right) \;\;, \nonumber\\ 
A_i^a & = & a \left[ \left( Q + \delta Q \right) \delta_{ai} + \partial_i \left( M_a + \partial_a M \right) 
+ \epsilon_{iab} \left( U_b + \partial_b U \right) + t_{ia} \right] \;\;, \nonumber\\
g_{00} & = & - a^2 \left( 1 - 2 \phi \right) \;\;, \nonumber\\
g_{0i} & = & a^2 \left( B_i + \partial_i B \right) \;\;, \nonumber\\
g_{ij} & = & a^2 \left[ \left( 1 + 2 \psi \right) \delta_{ij} + 2 \partial_i \partial_j E + \partial_i E_j + \partial_j E_i + h_{ij} \right]
\;\;,
\end{eqnarray}
where $a$ denotes both the SU(2) index and the scale factor (as we believe that this does not cause ambiguity), while $i=1,2,3$ ranges over the spatial coordinates. The modes $t_{ia}$ and $h_{ij}$ are transverse and traceless 
($\partial_i h_{ij} = \partial_i t_{ia} = \partial_a t_{ia} = t_{ii} = h_{ii} = 0$) and we denote them as ``tensor modes''. The modes $Y_a , M_a, U_a, B_i , E_i$ are transverse ($\partial_i Y_i = \dots = \partial_i E_i = 0$), and we denote them as ``vector modes''. We denote the remaining modes as ``scalar modes''. These three groups of modes are separate from each other at the linearized level, and can be studied independently  \cite{Dimastrogiovanni:2012ew}.~\footnote{The terms ``tensor/vector/scalar" are somewhat  misnomers for some of the modes, since they refer to the transformation properties of the corresponding modes in $\delta g$        under spatial rotations, but not to the transformation properties  of the modes in $\delta A$      (for instance, given that SU(2) indices have been used in that decomposition, the mode $M_a$ is not a vector under a spatial rotation).}
 
As shown in  \cite{Dimastrogiovanni:2012ew}, we can always choose the gauge $\psi=E=E_i=U=U_i=0$. This gauge choice completely fixes the freedom associated to general coordinate and SU(2) transformations. We Fourier transform each perturbation according to 
\begin{equation}
\delta \left( t , \vec{x} \right) = \int \frac{d^3 k}{\left( 2 \pi \right)^{3/2} } \, {\rm e}^{i \vec{k} \cdot \vec{x}} \, \delta \left( t , \vec{k} \right) \;\;.
\label{FT}
\end{equation}

We are only interested in studying the linearized theory of the perturbations. This amounts in expanding the equations of motion of the system up to first order in the perturbations, or in expanding the action of the system in quadratic order in the perturbation (this is equivalent, since the linearized equations for the perturbations are obtained by extremizing the quadratic action for the perturbations). In both these computations, the modes are decoupled from each other (since the perturbations are coupled to each other only at the nonlinear level). Therefore, without loss of generality in the linearized computation, we can choose the momentum $\vec{k}$ of the modes to be oriented along one given direction, 
which we  choose to be the $3$rd one~\footnote{Equivalently, starting from a mode with generic $\vec{k}$, we can always rotate the axis such that $k_x=k_y=0$ and simultaneously perform a SU(2) global transformation such that the background relation $\langle A_\mu^a \rangle \propto \delta_\mu^a$ is preserved. This also proves that we can set $k=k_z$ with no loss of generality at the linearized level \cite{Dimastrogiovanni:2012ew}.}

We end up with the following perturbations:
\begin{eqnarray}
{\rm scalar}: & &  \delta A_\mu^1 = a \left( 0 , \delta Q , 0 , 0 \right) \;\;,\;\;  
     \delta A_\mu^2 = a \left( 0 , 0 ,  \delta Q , 0   \right) \;\;,\;\;  
     \delta A_\mu^3 = a \left( \partial_z Y , 0 , 0 , \delta Q + \partial_z^2 M   \right) \;\;,  \nonumber\\
& & \delta g_{00} = a^2 2 \phi  \;\;,\;\; \delta g_{03} = a^2 \partial_z B  \;\;,  \nonumber\\
{\rm vector}: & &  \delta A_\mu^i = a \left( Y_i ,  0 , 0 , \partial_z M_i  \right) \;\;,\;\;
\delta g_{0i} = a^2 B_i \;\;,\;\; i=1,2 \nonumber\\
{\rm tensor}: & & \delta A_\mu^1 = a \left( 0 , t_+ , t_\times , 0 \right) \;\;,\;\; 
   \delta A_\mu^2 = a \left( 0 , t_\times , - t_+ , 0 \right) \;\;,\;\; 
  \delta g_{11} = - \delta g_{22} = a^2 h_+ \;\;,\;\; \delta g_{12} = a^2 h_\times \;\;.
\label{modes}
\end{eqnarray}

Namely, after $7$ perturbations are removed by the gauge fixing, we end up with $5$ modes in the scalar sector, $6$ modes in the vector sector, and $4$ modes in the tensor sector. Not all these modes  correspond to physically propagating independent degrees of freedom. Due to the structure of the kinetic terms, the modes originating from 
$\delta g_{0\mu}$ and $\delta A_0^a$ enter in the quadratic action of the perturbations without time derivatives (up to boundary terms) and are commonly denoted as ``non-dynamical'' perturbations.~\footnote{This justifies our gauge choice  \cite{Dimastrogiovanni:2012ew}, since this choice preserves all the non-dynamical  $\delta g_{0\mu}$ and $\delta A_0^a$ modes. The number of physically independent propagating degrees of freedom is gauge independent, and in other gauges the non-dynamical degrees of freedom appear as (in general, nontrivial) linear combinations of the modes preserved in that gauge.} Extremizing the quadratic action of the perturbations with respect to these modes provide equations that are algebraic in the non-dynamical modes. These equations are called ``constraint equations''. We solve these equations by expressing the non-dynamical modes in terms of the dynamical modes and their first time derivatives. 

Therefore, the non-dynamical modes do not introduce additional degrees of freedom in the initial condition, but are uniquely determined in terms of the dynamical modes. We thus see that Gauge-flation is characterized by $22$ (the starting number of modes) minus $7$ (gauge-fixing) minus $7$ (the amount of non-dynamical modes) equal $8$ physically propagating degrees of freedom. In our choice (\ref{modes}), these are the modes $\delta Q$ and  $M$ in the scalar sector, the modes $M_1$ and $M_2$ (in the vector sector), and all the $4$ modes in the tensor sector. 

In the remainder of this  work we disregard the vector sector, since it is possible to show that the model is ruled out by observations from the study of the scalar and tensor sectors alone.

\subsection{Formal expressions for the  the linearized equations, and formal solutions}
 
\label{subsec:formal-deco}

In this Subsection we study the quadratic action for the perturbations and the corresponding linearized equations of motion at the formal level. Particular care is taken in distinguishing the role of the non-dynamical vs the dynamical modes. The discussion follows and summarizes the analogous one presented in Appendix A of  \cite{Dimastrogiovanni:2012ew}. 

Once expanded at second order in the perturbations, the action of the system splits in three separate parts
\begin{equation}
S_{\rm quadratic} = S_{\rm scalar} + S_{\rm vector} + S_{\rm tensor} \;\;,
\end{equation}
all of which are Hermitian. The tensor action only contains dynamical modes, while the other two actions contain both dynamical and non-dynamical modes (see the final part of the previous Subsection). 

In complete generality, a Hermitian and quadratic action for a set of dynamical modes $\left\{ X_i \right\}$ and a set of non-dynamical modes $ \left\{ N_i \right\}$ is of the form (in matrix notation)  
\begin{equation}
S = \int d \tau d^3 k \left[ X^{'\dagger} A X' + \left(  X^{'\dagger} B X + {\rm h.c.} \right)
+ X^\dagger C X + \left( N^\dagger D X' + {\rm h.c.} \right) +  \left( N^\dagger E X + {\rm h.c.} \right) + N^\dagger F N \right]
\;\;,
\label{S2-formal}
\end{equation}
where $A,C,F$ are Hermitian, and (up to an integration by parts) $B$ is anti-Hermitian.  These matrices are function of background quantities, and therefore they are time dependent. Extremizing (\ref{S2-formal}) with respect to the non-dynamical and the dynamical   fields, we find, respectively,
\begin{eqnarray}
&& D X' + E X + F N = 0 \;\;, \nonumber\\
&& \left( A X' + B X + D^\dagger N \right)' - B^\dagger X' - C X - E^\dagger N = 0 \;\;.
\label{eom1-formal}
\end{eqnarray}
These equations are nothing but the linearized equations for the perturbations, and can  equivalently be obtained by expanding the exact SU(2) and Einstein equations of the model at first order in the perturbations. 

The first line in (\ref{eom1-formal}) is the constraint equations of the system, which  are solved by
\begin{equation}
N = - F^{-1} \left( D X' + E X \right) \;\;.
\label{constraint-sol}
\end{equation}
We can insert this solution back into the second line of (\ref{eom1-formal}), so that the remaining linearized equations
(that we still have to solve) are expressed in terms of dynamical variables only:
\begin{eqnarray}
&&
\left( A - D^\dagger F^{-1} D \right) X'' + \left[ \left( A-D^\dagger F^{-1} D \right)' + 
\left( B - D^\dagger F^{-1} E - {\rm h.c.} \right) \right] X'  \nonumber\\
& & \quad\quad\quad\quad  \quad\quad\quad\quad  \quad\quad\quad\quad  \quad\quad\quad\quad  + 
\left[ \left( B - D^\dagger F^{-1} E \right)' - C + E^\dagger F^{-1} E \right] X = 0 \;\;.
\label{eom2-formal}
\end{eqnarray} 
We can also insert the solution (\ref{constraint-sol}) back into the starting action (\ref{S2-formal}), and obtain a quadratic action for the dynamical variables only
\begin{equation}
S = \int d \tau d^3 k \left[ X^{'\dagger} \left( A - D^\dagger F^{-1} D \right) X' + 
\left( X^{'\dagger} \left( B - D^\dagger F^{-1} E \right) X + {\rm h.c.} \right)
+ X^\dagger \left( C - E^\dagger F^{-1} E \right) X \right] \;\;.
\label{S22-formal}
\end{equation}
Obviously, extremizing the quadratic action (\ref{S22-formal}) precisely produces the equations of motion (\ref{eom2-formal}).

To summarize, we have explicitly outlined at the formal level the procedure that is commonly denoted as ``integrating out the non-dynamical variables''. We have obtained a quadratic action in terms of the dynamical variables only. All the information necessary to solve the system of these variables is contained in this action, and only in it. As we have seen, this action provides the linearized equations  (\ref{eom2-formal})  in terms of the dynamical variables only; moreover,
as we discuss in Subsection \ref{subsec:PS}, this action is the starting point of the quantization of the perturbations, which will allow us to set the initial conditions for the modes. Therefore, the action  (\ref{S22-formal}) completely determines the Cauchy problem that uniquely determines the solutions for the dynamical modes. 

In our computations below we explicitly derive the action (\ref{S22-formal}) for the dynamical variables of Gauge-flation. We stress that, once the explicit form of  (\ref{S22-formal}) has been obtained, it contains all the information needed to derive the final solutions for the dynamical modes, and that the constraint equations can no longer be used for this purpose (we have just used them to eliminate the non-dynamical modes, and using them again would reintroduce the non-dynamical modes in the system that we are solving). In particular, we can no longer use the constraint equations to ``learn'' anything new on the dynamical variables beyond what the action   (\ref{S22-formal}) indicates. Only after the solution for the dynamical variables have been obtained from   (\ref{S22-formal}), the constraint equations can be used in the form  (\ref{constraint-sol}) to determine the explicit solution for the non-dynamical modes in terms of the explicit solutions for the dynamical modes that we have obtained from  (\ref{S22-formal}).

The procedure outlined in this Subsection  is commonly used in the case of scalar field inflation. For instance, in 
 appendix A of  \cite{Dimastrogiovanni:2012ew} we explicitly worked out the example of a single scalar field, and showed how the results of \cite{Mukhanov:1990me} are reproduced. The fact that some of the variables in the problem at hand originated from a vector multiplet does not impact this discussion in any way, since the study presented here 
 starts from the action (\ref{S2-formal}), which can always been written in this form for any system of perturbations.

\subsection{Quantization and power sectra}

\label{subsec:PS}

The kinetic matrix $A-D^\dagger F^{-1} D$ in  (\ref{S22-formal}) is Hermitian, and can be
diagonalized through
\begin{equation}
X_i = {\cal M}_{ij} \Delta_j \;\;.
\label{XtoD}
\end{equation}
The ``mix term'' in the resulting action  in terms of $\Delta$ can be  simplified by removing a boundary term, and we can write the action in the form 
\begin{equation}
S = \frac{1}{2} \int d \tau d^3 k \left[ \Delta^{' \dagger} T  \Delta' +  \Delta^{' \dagger}   K \Delta - \Delta^\dagger K \Delta' - \Delta^\dagger \Omega^2 \Delta \right] \;\;,\;\; T = 1 \;\;,
\label{S23-formal}
\end{equation}
where the Hermitianity of the action implies that  $K$ is anti-Hermitian, and $\Omega^2$ is Hermitian. 
For the model we are investigating, the matrices $K$ and $\Omega^2$  actually turn out to be real, and therefore we assume that this is the case also in the present discussion.

We choose to perform the quantization in terms of the fields $\Delta_i$:
\begin{equation}
\Delta_i \equiv {\cal D}_{ij} a_j +  {\cal D}_{ij}^* a_j^\dagger \;\;\;\;\;\;,\;\;\;\;\;\;
\left[ a_i \left( \vec{k} \right) , a_j ^\dagger \left( \vec{p} \right) \right] = \delta^{(3)} \left( \vec{k} - \vec{p} \right) \delta_{ij} \;\;.
\label{D-deco}
\end{equation}
To perform the quantization, besides these relations, we  need to impose the equal time commutation relations (ETCR) between $\Delta_i$ and its conjugate momentum in (\ref{S23-formal}):  
\begin{equation}
\left[ \Delta_i \left( t , \vec{x} \right) , \Pi_j \left( t ,  \vec{y} \right) \right] = i \delta_{ij} \delta^{(3) } \left( \vec{x} - \vec{y} \right) 
 \;\;\;\;\;\;,\;\;\;\;\;\; 
\Pi_i \equiv \frac{\partial L}{\partial \Delta_i'}  \;\;, 
\label{conjugate}
\end{equation}
where, with an abuse of notation, in this relation (and only in this relation)  the fields are in real space. Decomposing $\Pi$ in terms of the same annihilation / creation operators as in (\ref{D-deco}), we have 
\begin{equation}
\Pi_i  =    \pi_{ij} a_j + \pi_{ij}^* a_j^\dagger \;\;\;,\;\;\;  \pi_{ij} = {\cal D}_{ij}' + K_{il} {\cal D}_{lj} \;\;.
\end{equation}

The relations (\ref{D-deco}) and (\ref{conjugate}) can be simultaneously imposed only if    the condition,
\begin{equation}
\left[   {\cal D} \pi^\dagger - {\cal D}^* \pi^T \right]_{ij} = i \, \delta_{ij} \;\;,
\label{wronskian}
\end{equation}
is satisfied. Notice that     we can impose (\ref{wronskian}) as an initial condition; however, consistency of the computation requires that this condition holds at all times.   This is the case if the initial conditions also satisfy  
\begin{equation}
\pi \pi^\dagger - \pi^* \pi^T = {\cal D} {\cal D}^\dagger - {\cal D}^* {\cal D}^T = 0 \;\;,
\label{wronskian2}
\end{equation}
namely if    the products $\pi \pi^\dagger $ and ${\cal D} {\cal D}^\dagger$ are real. Indeed,  the three conditions in (\ref{wronskian}) and (\ref{wronskian2}) are preserved by the equations of motion if they all hold initially: 
to verify this, one can write   the equations of motion following from  (\ref{S23-formal})  in terms of ${\cal D}$ and $\pi$. Using these equations, one can then show that the first derivatives of (\ref{wronskian}) and (\ref{wronskian2}) vanish if these relations hold. Therefore, if these relations hold initially, they are preserved by the evolution, and continue to hold at all times. 

The condition (\ref{wronskian}) generalizes to the multi-field case the standard Wronskian condition imposed by   the single field quantization. The conditions (\ref{wronskian2}) instead vanish identically in the single field case, and have no counterpart in this  case. We will collectively denote the three conditions  (\ref{wronskian}) and  (\ref{wronskian2}) 
 as Wronskian conditions in the remainder of this work.

In the explicit computations below, we show that, in the initial sub-horizon regime, the modes  can be chosen according to the positive frequency initial adiabatic vacuum prescription, and satisfying the conditions  (\ref{wronskian}) and (\ref{wronskian2}). Starting from these initial conditions, we then compute the time evolution for the mode functions and obtain the solution at late time, when the mode is outside the horizon. The solutions provide the late time correlators of any observable of our interest. Specifically, the observables we are interested in can be written as linear combinations of the dynamical variables and their time derivatives:
\begin{equation}
{\cal O } \left( t , \vec{x} \right) = \int \frac{d^3 k}{\left( 2 \pi \right)^{3/2} } \, {\rm e}^{i \vec{k} \cdot \vec{x}} \, 
{\cal O } \left( t , \vec{k} \right) \;\;,\;\;
{\cal O } \left( t , \vec{k} \right) =  c_i \left( t ,\,  \vec{k} \right) X_i \left( t ,\, \vec{k} \right) +  d_i \left( t ,\,  \vec{k} \right) X_i' \left( t ,\, \vec{k} \right) 
 \;\;.
\label{observable}
\end{equation}
where the reality of ${\cal O}$ enforces that  ${\cal O }^* \left( \vec{k} \right) = {\cal O} \left( - \vec{k} \right)$. We can therefore decompose also ${\cal O}$ in terms of annihilation / creation operators
\begin{eqnarray}
{\cal O } & = & {\cal O}_i \left( \vec{k} \right) a_i \left( \vec{k} \right) + {\cal O}^*_i  \left( -  \vec{k} \right) a_i \left( -  \vec{k} \right)^\dagger \;\;, \nonumber\\
{\cal O}_i & = & c_j {\cal M}_{jl} {\cal D}_{li} + d_j \left( {\cal M}_{jl} {\cal D}_{li} \right)' \;\;.
\label{cal-O}
\end{eqnarray}
and we obtain the correlator
\begin{equation}
\langle {\cal O } \left( t , \vec{x} \right)  {\cal O } \left( t , \vec{y} \right) \rangle = \int \frac{d^3 k}{\left( 2 \pi \right)^3} 
{\rm e}^{i \vec{k} \cdot \left( \vec{x} - \vec{y} \right)} \, \sum_i  \left\vert {\cal O}_i \left( \vec{k} \right) \right\vert^2
\end{equation}

Assuming statistical isotropy, $ \left\vert {\cal O}_i \left( \vec{k} \right) \right\vert^2 =  \left\vert {\cal O}_i \left( k \right) \right\vert^2$, and the correlator can be written in terms of the isotropic power spectrum
\begin{eqnarray}
& & 
\langle {\cal O } \left( t , \vec{x} \right)  {\cal O } \left( t , \vec{y} \right) \rangle = \int \frac{d k}{k} \, \frac{\sin \left( k \, r \right)}{k \, r} \, P_{{\cal O}{\cal O}} \left( k \right) \;\;\;,\;\;\; r \equiv \vert \vec{x} - \vec{y} \vert  \;\;,  \nonumber\\
& & 
P_{{\cal O}{\cal O}} \left( k \right) = \frac{k^3}{2 \pi^2} \sum_i \vert  {\cal O}_i \vert^2
\label{PS-OO}
\end{eqnarray}

We conclude this Subsection by noting a symmetry of the above solutions. Specifically, if we replace
\begin{equation}
{\cal D} \rightarrow {\cal D} \, U \;\;,
\label{phase-arbit}
\end{equation}
where $U$ is a constant and unitary matrix, and we perform the same transformation on ${\cal D}'$ and ${\cal \pi}$, the  conditions (\ref{wronskian}) and (\ref{wronskian2}), as well as the final power spectrum (\ref{PS-OO}) are unchanged. 
The matrix $U$ is arbitrary and  unphysical; this  generalizes to the $N$-field case the usual phase arbitrarily of the mode function in the single field case. We use this arbitrariness in Subsection  \ref{subsec:initial} to simplify the initial conditions.

\section{Tensor modes}

\label{sec:tensor}

The tensor sector of Gauge-flation is given in eq. (\ref{modes}). It is actually convenient to work in terms of  the left-handed and right-handed canonical modes, related to the fields given in  (\ref{modes})   through 
\begin{equation}
h_+ = \frac{h_L + h_R}{\sqrt{2}} \;\;,\;\;
h_\times = \frac{h_L - h_R}{i \sqrt{2}} \;\;,\;\;
t_+ = \frac{t_L + t_R}{\sqrt{2}} \;\;,\;\;
t_\times = \frac{t_L-t_R}{ i \sqrt{2}} \;\;.
\label{LR-def}
\end{equation}

We insert these modes in the action of the model, and expand to quadratic order. The action splits in two decoupled parts,
\begin{equation}
S_{\rm tensor } = S_L + S_R \;\;,
\end{equation}
so that the doublet $\left\{ h_L ,\, t_L \right\}$ is the counterpart of $\left\{ X \right\}_i$  - defined above eq. (\ref{S2-formal}) - in the left-handed sector (and analogously for the right-handed sector). We recall that there are no non-dynamical tensor modes. 

The modes in (\ref{LR-def}) are not canonically normalized. Canonical normalization is achieved through
\begin{equation}
h_L = \frac{\sqrt{2}}{M_p a} \, H_L \;\; , \;\; 
t_L = \frac{1}{\sqrt{2} a}  \, T_L
\label{canonical-tensor}
\end{equation}
(and analogously in the right-handed sector). This diagonalization is the explicit expression for (\ref{XtoD}), and leads to 
an action of the form
\begin{equation}
S_L = \frac{1}{2} \int d \tau d^3 k \left[ \Delta_L^{'\dagger}  \Delta_L' 
+  \Delta_L^{'\dagger} K_L  \Delta_L 
-  \Delta_L^\dagger K_L  \Delta_L' 
-   \Delta_L^\dagger \Omega_L^2  \Delta_L \right] \;\;,\;\;
\Delta_L = \left( \begin{array}{c} H_L \\ T_L \end{array} \right) \;\;, 
\label{tensor-S}
\end{equation}
and identically for the right-handed sector. The matrix $K_L$ is anti-symmetric, with
\begin{equation}
K_{L,12} = \frac{1}{M_p} \left( Q' + \frac{a'}{a} Q \right) \;\;,
\label{KL}
\end{equation}
while the $\Omega_L^2$ matrix is symmetric, with
\begin{eqnarray}
\Omega_{L,11}^2 & = & k^2 - 2 \frac{a'^2}{a^2} + \frac{3 g^2 a^2 Q^4}{M_p^2} - \frac{\left( a Q \right)^{'2}}{M_p^2 a^2} \;\;, \nonumber\\ 
\Omega_{L,12}^2 & = & k \frac{2 g a Q^2}{M_p} + \frac{\left( a Q \right)'}{a M_p} \frac{a'}{a} - \frac{2 \kappa g^2 Q^3}{M_p a^2} \frac{g^2 a^4 Q^4 + a^{'2} Q^2 - a^2 Q^{'2}}{1+\kappa g^2 Q^4} \;\;, \nonumber\\ 
\Omega_{L,22}^2 & = & k^2 - 2 k g a Q \left[ 1 + \kappa \frac{g^2 a^4 Q^4 + a^{' 2} Q^2 - a^2 Q^{' 2} }{a^4 \left( 1 + \kappa g^2 Q^4 \right) } \right]  + \frac{2 \kappa g^2 Q^2}{ a^2} \frac{g^2 a^4 Q^4 + a^{'2} Q^2 - a^2 Q^{'2}}{1+\kappa g^2 Q^4} \;\;. 
\label{omegaL}
\end{eqnarray}
The matrices $K_R$ and $\Omega_R^2$  are obtained from  $K_L$ and $\Omega_L^2$,  respectively,  by replacing $k \rightarrow - k$. This causes a difference between the two helicities, signaling a violation of parity invariance in the tensor sector, analogously to what happens for the model of Chromo-natural inflation \cite{Adshead:2013qp}.

To gain an analytical understanding of the solutions in the tensor sector, we first of all rewrite the above matrix elements eliminating as many parameters as possible (using some background relations), and then we approximate them in slow roll. Specifically, we first of all perform the following substitutions on each matrix element:
 \begin{eqnarray}
&& Q' \rightarrow - a Q H \delta \;\;,\;\; a' \rightarrow a^2 H \;\;,\;\; k \rightarrow p \,  a \;\;,\;\;
 \kappa \rightarrow   \frac{1}{H^2 \gamma Q^2} \;  \frac{\left( 1 - \delta \right)^2 + \gamma}{\left( 1 - \delta \right)^2} \, \frac{2-\epsilon}{\epsilon} \;\;, \nonumber\\ 
  && M_p \rightarrow Q \sqrt{\frac{\left( 1 - \delta \right)^2 + \gamma}{\epsilon}}
 \;\;,\;\; g \rightarrow \sqrt{\gamma} \frac{H}{Q} \;\;.
  \label{subs-before-slowroll}
  \end{eqnarray}
Even if they employ slow roll parameters, all these substitutions are exact. The first substitution follows from the definition of $\delta$ in eq.  (\ref{s2seven}), the second and third substitutions are standard relations between comoving/conformal and physical quantities ($p$ is the physical momentum of a mode),  while the last three substitutions follow directly from eqs.  (\ref{relation-kappa}), (\ref{relation-Mp}), and (\ref{def-gamma}), respectively.
From these substitutions, we obtain:
\begin{eqnarray}
\frac{ K_{L,12} }{a} & = & H \frac{\left( 1 - \delta \right) \sqrt{\epsilon}}{\sqrt{\gamma+\left( 1 - \delta \right)^2} } \;\;, \nonumber\\ 
\frac{ \Omega_{L,11}^2 }{a^2} & = & p^2 - H^2 \frac{\left( 1 - \delta \right)^2 \left( 2 + \epsilon \right) + \gamma \left( 2 - 3 \epsilon \right)}{\gamma + \left( 1 - \delta \right)^2} \;\; , \nonumber\\  
\frac{ \Omega_{L,12}^2 }{a^2} & = &   \frac{ H p \, 2 \sqrt{\gamma \epsilon}}{\sqrt{\gamma + \left( 1 - \delta \right)^2} } -
 \frac{ H^2  \sqrt{\epsilon}}{\sqrt{\gamma+\left( 1 - \delta \right)^2} } \frac{2 \gamma^2 \left( 2 - \epsilon \right) + 3 \gamma \left( 1 - \delta \right) \left( 2 - \epsilon \right) + 2 \left( 1 - \delta \right)^3 \left[ 1 + \delta \left( 2 - \epsilon \right) - \epsilon \right]}{2 \left( 1 - \delta \right)^2 + \gamma \left( 2 - \epsilon \right) } \;\; ,  \nonumber\\ 
\frac{ \Omega_{L,22}^2 }{a^2} & = &  p^2 - H p \; \frac{2  }{\sqrt{\gamma}} \frac{2 \gamma^2 \left( 2 - \epsilon \right) + \left( 1 - \delta \right)^3 \left( 1 + \delta \right) \left( 2 - \epsilon \right) + 2 \gamma \left( 1 - \delta \right) \left( 3 - \delta - \epsilon \right)}{2 \left( 1 - \delta \right)^3 + \gamma \left( 2 - \epsilon \right)}    \nonumber\\
& & \quad \quad 
+ H^2 \; 2 \left( 2 - \epsilon \right)  \frac{\gamma^2 + 2 \gamma \left( 1 - \delta \right) + \left( 1 - \delta \right)^3 \left( 1 + \delta \right)}{2 \left( 1 - \delta \right)^2 + \gamma \left( 2 - \epsilon \right) } \;\;.
\label{tensor-substituted}
\end{eqnarray}
We stress that these expressions are completely equivalent to the corresponding ones in eqs. (\ref{KL}) and  (\ref{omegaL}) and that no slow roll approximation has yet been done. However, the expressions in (\ref{tensor-substituted}) are much more transparent, as the only dimensional parameters are the physical momentum $p$ and he physical Hubble rate  $H$, while any explicit reference to $Q$ has been eliminated. Besides $H$ and $p$, the expressions (\ref{tensor-substituted}) are given in terms of three dimensionless parameters: the two slow-roll parameters $\epsilon$ and $\delta$, and the parameter $\gamma$.

All expressions in (\ref{tensor-substituted}) have a power law dependence on $p$ and $H$   of the type $K_{12} \propto  c_1 \, H \;,\; \Omega_{ij}^2 = c_2 p^2 + c_3 p H + c_4 H^2$.  Therefore these expressions immediately indicate which terms dominate in the sub-horizon and in the super-horizon regime. All the $c_i$ coefficients  are slowly evolving, and can be immediately expanded in slow roll. To perform the expansion, we
\begin{equation}
{\rm Substitute \;\; } \delta \rightarrow \frac{\gamma}{6 \left( 1 + \gamma \right)} \epsilon^2 \;\;\;\; {\rm and \; expand \; in \;\; } \epsilon \ll 1 \;\;,  
 \label{subs-slowroll}
 \end{equation}  
where the first expression is the slow roll solution for $\delta$, see eq. (\ref{sl2}). The expansion leads to 
 \begin{eqnarray}
\frac{ K_{L,12} }{a} & \simeq & H \frac{\sqrt{\epsilon}}{\sqrt{1+\gamma}} \;\;, \nonumber\\ 
\frac{ \Omega_{L,11}^2 }{a^2} & \simeq & p^2 - 2 H^2 \;\; ,  \nonumber\\
\frac{ \Omega_{L,12}^2 }{a^2} & \simeq & \frac{2 \sqrt{\gamma \epsilon}}{\sqrt{1+\gamma}} H p - \frac{\left( 1 + 2 \gamma \right) \sqrt{\epsilon} }{\sqrt{1+\gamma}} H^2 \;\; , \nonumber\\ 
\frac{ \Omega_{L,22}^2 }{a^2} & \simeq & p^2 - \frac{ 2  \left( 1 + 2 \gamma \right)}{\sqrt{\gamma}} \, p \,  H + 2 \left( 1 + \gamma \right) \, H^2 \;\;. 
\label{tensor-slow}
\end{eqnarray}
We stress that, as we do not disregard any of the $c_i$ coefficients, but we simply expand them in slow roll, the expressions in (\ref{tensor-slow}) provide a very accurate approximation to the exact expressions (\ref{tensor-substituted}) at all times (namely, for any value of $p/H$).   All of them agree with the exact expressions at all times, with an   ${\rm O } \left( \epsilon \right)$ accuracy or better.

From the expressions (\ref{tensor-slow}) we can gain an intuitive understanding of the tensor sector. We first of all note that, if it was not for the coupling with the $t_{L/R}$ modes from the gauge field,    the tensor mode helicities would obey the standard relation $h_{L/R}'' + \left( k^2 - 2 H^2 a^2 \right) h_{L/R} \simeq 0$. However, for some choice of parameters the mode $t_L$ exhibits a large tachyonic growth for some time close to horizon crossing, and it can then source a large growth of $h_L$. The same does not occur in the right-handed sector. The situation is completely analogous to what takes place in Chromo-natural inflation  \cite{Adshead:2013qp}.

To understand why $t_L$ can grow for some choice of parameters, and why the same does not take place for $t_R$, we can disregard their coupling to the $h_{L/R}$ modes. This is an accurate approximation, since the coupling affects $t_{L/R}$ through the $K_{12}$ and $\Omega_{12}^2$ terms, that are slow roll suppressed with respect to the $\Omega_{22}^2$ term at all times.   We see that
\begin{equation}
 \Omega_{L,22}^2 < 0 \;\;\;\; {\rm for } \;\;\; r_* - \Delta r < \frac{p}{H} < r_* + \Delta r \;\; , \;\;  
{\rm with } \;\;\;\;   r_* \equiv \frac{1+2 \gamma}{\sqrt{\gamma}} \;\;\; , \;\;\; 
\Delta r \equiv \frac{\sqrt{1+2\gamma+2\gamma^2}}{\sqrt{\gamma} } \;\; .
\end{equation}
Therefore, for any choice of $\gamma$, each mode of $t_L$ experiences a tachyonic instability in a neighborhood of $p/H = r_*$ (therefore, the tachyonic growth is absent at sufficiently small and sufficiently large scale, but always takes place close to horizon crossing). We recall that $ \Omega_{R,22}^2 $ is related to  $ \Omega_{R,22}^2 $ by $p \rightarrow - p$, so that  $ \Omega_{R,22}^2 $ is instead positive at all times, and there is no tachyonic growth for the right-handed modes. 
In the left-handed sector, for any fixed value of $H$, the most negative value of  $\Omega_{L,22}^2$ takes place at
\begin{equation}
\Omega_{L,22}^2 \Big\vert_{p=\frac{1+2 \gamma}{\sqrt{\gamma}} H } \simeq - H^2 \;  \frac{1+2\gamma+2\gamma^2}{\gamma} \;\;. 
\end{equation}

Both the minimum value of $\Omega_{L,22}^2 $ and the duration of the tachyonic phase increase both at large and small $\gamma$. However, as we shall see in the next Section, the region $\gamma < 2$ is excluded  as the scalar perturbations are unstable there. For $\gamma > 2$, we see that the instability grows with $\gamma$, and so we should expect that too large values of $\gamma$ are excluded because the growth of $t_L$ will source too large a growth of $h_L$. In Section \ref{sec:pheno} we show that this is indeed the case, and that the Planck constraint $r \lta 0.11$ \cite{Ade:2013zuv} forces $\gamma \lta 5$ (the precise value depending on the number of e-folds of inflation).

 The actions $S_{L/R} \rightarrow \frac{1}{2} \int d \tau d^3 k \left[ \vert \Delta_{L/R,i}' \vert^2 - k^2 \vert \Delta_{L/R,i} \vert^2 \right]$ at asymptotically early times. This gives the initial conditions 
\begin{equation}
\sqrt{2 k } {\cal D}_{L,{\rm in}} = \mathbb{I} \;\; , \;\; 
\sqrt{2 k } {\cal D}_{L,{\rm in}}' = - i k \,   \mathbb{I} \;\; ,  
\end{equation}
where $ \mathbb{I} $ is the identity operator, $ \mathbb{I}_{ij} = \delta_{ij} $.

Starting from these initial conditions, we numerically integrate  the equations following from (\ref{tensor-S}) and (\ref{D-deco}):
\begin{equation}
{\tilde \partial }^2 \left[ \sqrt{2 k} {\cal D}_L \right] + 
\left[ 2 \frac{\sqrt{\kappa } M_p K_L}{a} + \frac{{\tilde \partial } a}{a}  \right] {\tilde \partial }  \left[ \sqrt{2 k} {\cal D}_L \right] 
+ \left[ \frac{ {\tilde \partial } \left( \sqrt{\kappa} M_p K_L \right) }{a} + \frac{\kappa M_p^2 \Omega_L^2}{a^2} \right]  \left[ \sqrt{2 k} {\cal D}_L \right] = 0   \;\;.
\label{eom-DL}
\end{equation}
(and identically for the right-handed sector) and obtain the power 
\begin{equation}
P_{L/R} = \frac{k^3}{2 \pi^2} \frac{2}{M_p^2 a^2} \frac{1}{2 k} \sum_{i=1}^2  \left\vert \sqrt{2 k} {\cal D}_{L/R,1i} \right\vert^2 =  \frac{1}{\kappa M_p^4}  \;  \frac{ {\tilde p}^2 }{2 \pi^2}    \sum_{i=1}^2  \left\vert \sqrt{2 k} {\cal D}_{L/R,1i} \right\vert^2  \;\;,
\label{PLR}
\end{equation}
where in the second expression we defined the dimensionless physical momentum 
\begin{equation}
{\tilde p} \equiv \sqrt{\kappa} M_p p \;\; .
\label{p-tilde} 
 \end{equation}

Eq. (\ref{eom-DL}) has been written in terms of the dimensionless quantities that we use in our numerical integration:
namely, the quantities defined in (\ref{tilde-combo}) and ${\tilde p}$. We recall that  ${\tilde \partial}$ denotes derivative with respect to the dimensionless physical time ${\tilde t}$, related to the derivative with respect to conformal time $\tau$ by $\frac{\partial }{\partial \tau } = \frac{a}{\sqrt{\kappa} M_p  } \, {\tilde \partial} $. In this way, all the matrix elements appearing in (\ref{eom-DL}) can be expressed solely in terms of dimensionless quantities, and are ``ready'' for the numerical integration. For instance,
\begin{equation}
\frac{\sqrt{\kappa }  M_p K_{L,12}}{a} = {\tilde \partial } {\tilde Q } + \frac{{\tilde \partial } a}{a} {\tilde Q} \;\; , 
\end{equation}
and analogously for all the other matrix elements. We recall that   $ \frac{{\tilde \partial } a}{a} $ is the Hubble rate in rescaled physical time.

\begin{figure}
\centering
\includegraphics[width=0.45\textwidth]{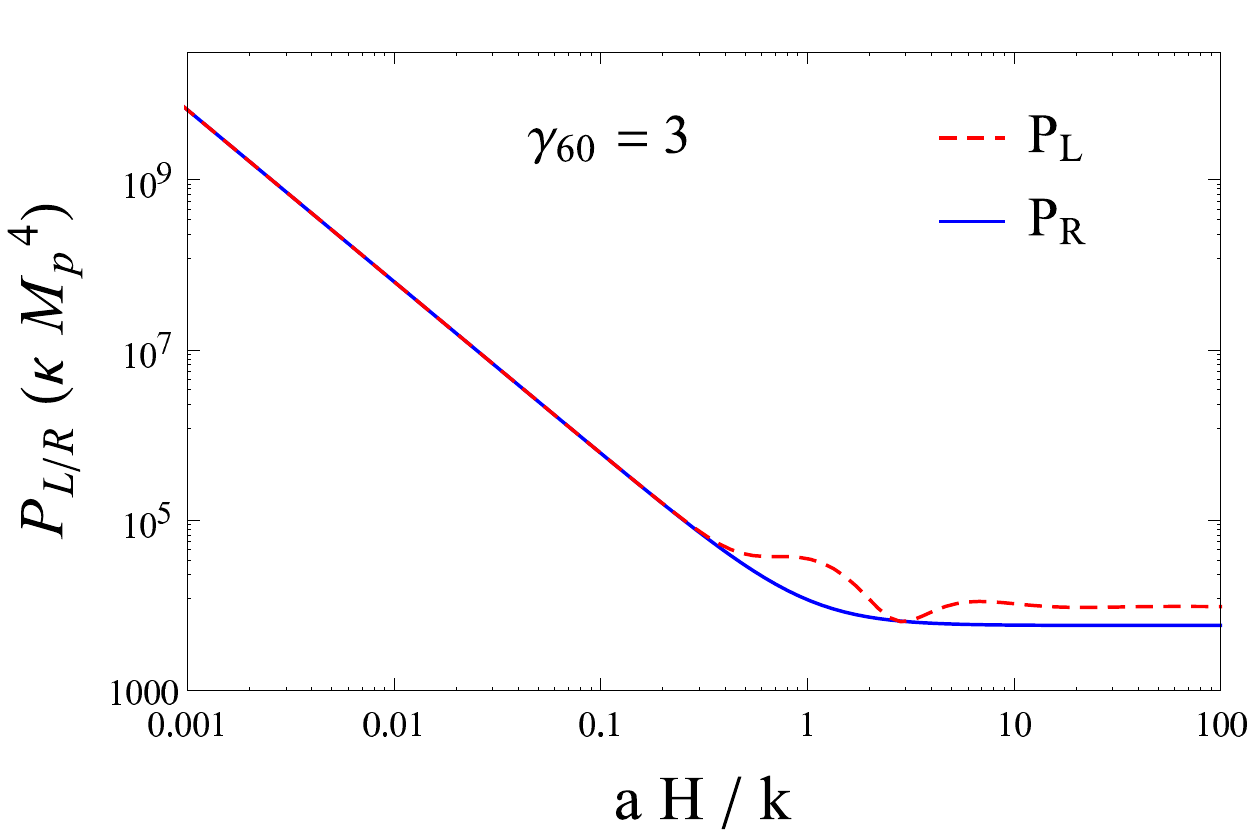}
\includegraphics[width=0.45\textwidth]{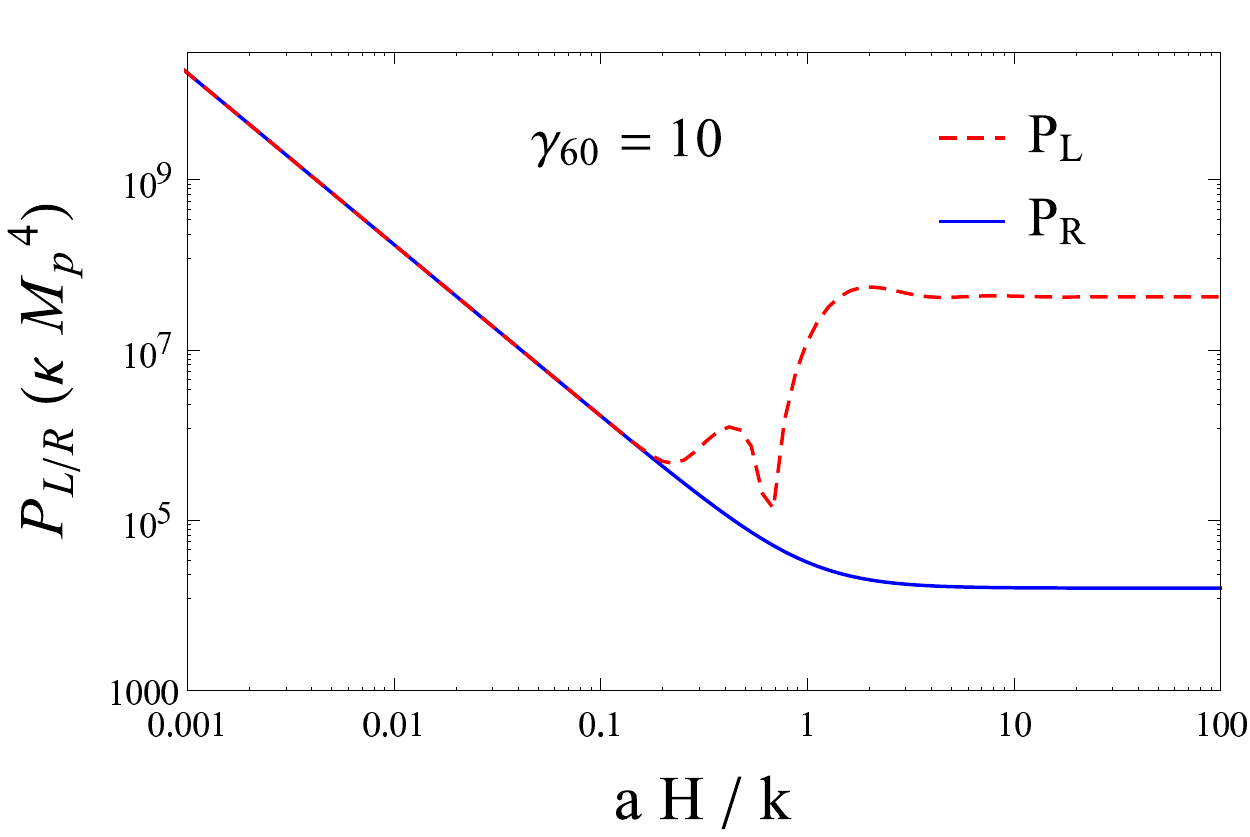}
\caption{ Growth of the power and freeze out for the left-handed and right-handed gravity waves, for two different values of $\gamma_{60} = 3$ (left panel) and for  $\gamma_{60} = 10$ (right panel),   where $\gamma_{60}$ is the value 
 $\gamma_{\rm in}$ at $60$ e-folds before the end of inflation. Both panel show the evolution of the power for a mode that leaves the horizon $60$ e-folds before the end of inflation. The $x-$axis is the  ratio between the Hubble horizon and the wavelength of the mode, and therefore it is a measure of time. Horizon crossing occurs at $a H / k = 1$.}     
\label{fig:PLPR}
\end{figure}

In Figure \ref{fig:PLPR}, we show the time evolution of the power $P_L$ and $P_R$  for a mode that leaves the horizon at $60$ e-folds before the end of inflation. In the left panel and right panel we show the evolution for $\gamma_{\rm in} = 3$ and $\gamma_{\rm in} = 10$, respectively. In both cases, the initial values of the two powers coincide at early times (since the early time actions in the left-handed and right-handed sectors are identical), but then $P_L > P_R$ at late times, signaling the breaking of parity in the tensor sector, and the tachyonic growth in the left-handed sector that we have also observed analytically. As discussed above, the final value of $P_L$ grows with growing $\gamma$.

\section{Scalar modes}

\label{sec:scalar}

In this Section we study the scalar system of perturbations of Gauge-flation in the gauge (\ref{modes}). The discussion is divided in several parts. In Subsection \ref{subsec:S2scal-explicit} we provide the    explicit form of the quadratic action of the perturbations. In Subsection \ref{subsec:initial} we provide the initial conditions imposed by this action and by the conditions (\ref{wronskian}) and (\ref{wronskian2}).      In Subsection \ref{subsec:scalarsol} we study the numerical evolution of the equations of motion following from this action, and we present some examples.

\subsection{The quadratic action}

\label{subsec:S2scal-explicit}

Introducing the scalar modes in (\ref{modes}) into the action of  Gauge-flation, and expanding up to second order, we obtain 
\begin{eqnarray}
 S_{\rm scalar}  & = & \int d \tau d^3 k \, {\cal L}_{\rm scalar} \;\;, \nonumber\\
  {\cal L}_{\rm scalar} & = &  
 \frac{a^2 \left( 1 + g^2 \kappa Q^4 \right)}{6}    \left\vert 3 \delta Q' - k^2 M' \right\vert^2 + \frac{k^4 a^2}{3}  \left\vert M' \right\vert^2 - \frac{a^2}{3} \Bigg\{ \frac{k^2}{3} + \frac{g^2 \kappa Q^4\left( 1 + g^2 \kappa Q^4 \right)  }{2 a^2 M_p^2 } \left( a Q \right)^{' 2}  \nonumber\\
& & \!\!\!\! \!\!\!\!  \!\!\!\!  \!\!\!\!  \!\!\!\!  \!\!\!\!   \!\!\!\!  \!\!\!\!   \!\!\!\! 
 + \frac{1}{1+g^2 \kappa Q^4} \left[ g^2 Q^2 \left( 3 - g^2 \kappa Q^4 \right) \left( a^2 + \kappa Q^2 \frac{a'^2}{a^2} + \kappa Q^{'2} \right) - 8 g^2 \kappa Q^4  \frac{a'^2}{a^2} \right] \Bigg\} \left\vert 3 \delta Q - k^2 M \right\vert^2 
  \nonumber\\
& &  \!\!\!\! \!\!\!\!  \!\!\!\!  \!\!\!\!  \!\!\!\!  \!\!\!\!   \!\!\!\!  \!\!\!\!   \!\!\!\! 
  - \frac{k^4 a^2}{3} \left( \delta Q^* M + {\rm h.c.} \right)
 + \frac{a^2 k^4}{3}  \left[ \frac{k^2}{3} - \frac{g^2 \kappa Q^4}{a^2 M_p^2 } \left( a Q \right)^{' 2}
- \frac{2 g^2 \kappa Q^2}{1+g^2 \kappa Q^4} \left( g^2 a^2 Q^4 + \frac{a^{'2}}{a^2} Q^2 - Q^{'2} \right) \right] \left\vert M \right\vert^2 \nonumber\\
& &  \!\!\!\!  \!\!\!\!  \!\!\!\!  \!\!\!\!  \!\!\!\!  \!\!\!\!   \!\!\!\!  \!\!\!\!   \!\!\!\! 
 + \Bigg\{  \frac{a^2 k^2}{6} \Bigg[  \left( 1 + g^2 \kappa Q^4 \right) \left( 3 \delta Q^{'*} - k^2 M^{'*} \right) - 2  k^2 M^{'*}  + \frac{1}{a} \left[ \left( 1 + g^2 \kappa Q^4 \right) a' + 2 g^2 \kappa Q^3 \left( a Q \right)' \right] \left( 3 \delta Q^* - k^2 M^* \right) 
  \nonumber\\
& &  \!\!\!\!  \!\!\!\!  \!\!\!\!  \!\!\!\!  \!\!\!\!  \!\!\!\!   \!\!\!\!  \!\!\!\!   \!\!\!\! 
 - 2 k^2 \frac{a'}{a} M^* \Bigg] Y + {\rm h. c. } \Bigg\} + k^2 a^2 \left[ g^2 a^2 Q^2 + \frac{k^2}{6} \left( 3 + g^2 \kappa Q^4 \right) \right] \left\vert Y \right\vert^2
+ \frac{a}{2} \left( 1 + g^2 \kappa Q^4 \right) \left( a Q \right)' \left[ \phi^* \left( 3 \delta Q' - k^2 M' \right) + {\rm h. c. } \right]
\nonumber\\ 
& &  \!\!\!\!  \!\!\!\!  \!\!\!\!  \!\!\!\!  \!\!\!\!  \!\!\!\!   \!\!\!\!  \!\!\!\!   \!\!\!\! 
+ \left[  \frac{a' \left( 1 + g^2 \kappa Q^4 \right)}{2} \left( a Q \right)' + g^2 \kappa Q^3 \left( a Q \right)^{' 2} + g^2 a^4 Q^3 \right] \left[ \phi^*  \left( 3 \delta Q - k^2 M \right) + {\rm h. c. } \right]
+ \frac{a k^2 \left( 1 + g^2 \kappa Q^4 \right) }{2} \left( a Q \right)' \left( Y^* \phi +  {\rm h.c.} \right) \nonumber\\
& &  \!\!\!\!  \!\!\!\!  \!\!\!\!  \!\!\!\!  \!\!\!\!  \!\!\!\!   \!\!\!\!  \!\!\!\!   \!\!\!\! 
- a k^2 \left( a Q \right)' \left[ B^* \delta Q + {\rm h. c. } \right] - g^2 k^2 a^4 Q^3 \left( Y^* B + {\rm h.c.} \right)
+ \frac{3}{4} \left[ 3 \left( 1 + g^2 \kappa Q^4 \right) \left( a Q \right)^{'2} + g^2 a^4 Q^4 - 6 M_p^2 a^{'2} \right] \left\vert \phi \right\vert^2 \nonumber\\
& &  \!\!\!\!  \!\!\!\!  \!\!\!\!  \!\!\!\!  \!\!\!\!  \!\!\!\!   \!\!\!\!  \!\!\!\!   \!\!\!\! 
+ \frac{k^2}{4} \left[ g^2 a^4 Q^4 + 6 M_p^2 a^{'2} - 3 \left( 1 + g^2 \kappa Q^4 \right) \left( a Q \right)^{'2} \right] \left\vert B \right\vert^2 - k^2 M_p^2 a a' \left( B^* \phi + {\rm h. c. } \right) \;\;.
\label{s2sca-explicit}
\end{eqnarray}
which is indeed of the form (\ref{S2-formal}) in terms of the dynamical modes $X_i = \left\{ \delta Q , M \right\}_i$ and of the non-dynamical modes $N_i = \left\{ Y , \phi , B \right\}_i$. We integrated out the non-dynamical modes as outlined in Subsection \ref{subsec:formal-deco}, leading to an action of the form  (\ref{S22-formal}) in terms of the dynamical modes only. 

We transformed the two dynamical fields according to eq. (\ref{XtoD}), with
\begin{equation}
X_i = {\cal M}_{ij} \Delta_j \;\;\;,\;\;\; 
{\cal M} = \left( \begin{array}{cc}
- \frac{1}{\sqrt{6} a } \sqrt{ 1 + \frac{2}{1+\kappa g^2 Q^4}} & 0 \\
- \sqrt{\frac{3}{2}} \frac{1}{a k^2} \sqrt{1-\frac{2}{3+\kappa g^2 Q^4}} & 
- \frac{1}{a k^2} \sqrt{\frac{k^2}{2 a^2 g^2 Q^2 } + \frac{3}{3+\kappa g^2 Q^4}}
\end{array} \right) \;\;.
\label{XtoD-explicit}
\end{equation}
The resulting action is of the  form
\begin{equation}
S = \frac{1}{2} \int d \tau d^3 k \left[ \Delta^{' \dagger} T  \Delta' +  \Delta^{' \dagger}   K \Delta - \Delta^\dagger K \Delta' - \Delta^\dagger \Omega^2 \Delta \right] \;\;,\;\; 
\label{S24-formal}
\end{equation}
which is nearly of the form (\ref{S23-formal}), with the difference that the matrix $T$ is not the identity in this expression.

Formally speaking, it is straightforward to set the matrix $T$ equal to the identity for an action of the type (\ref{S24-formal}), 
through a second field redefinition, and to include this second  redefinition   in the transformation matrix ${\cal M}$. However, the explicit form of the matrix ${\cal M}$ would be extremely involved for the problem at hand. Fortunately, this additional step is not necessary. Indeed, the action 
(\ref{S24-formal}) is needed for two purposes. The first purpose is to obtain the evolution equations for the mode functions. Performing the decomposition (\ref{D-deco}), the equations of motion following from (\ref{S24-formal}) are
\begin{eqnarray}
&& {\cal D}'' + \alpha \, {\cal D}' + \beta \, {\cal D} = 0 \;\;, \nonumber\\
&&
\alpha  \equiv   T^{-1} \left( T' + 2 K \right)  \;\;,\;\;  
\beta  \equiv   T^{-1} \left( K' + \Omega^2 \right)  \;\;, 
\label{eom-formal}
\end{eqnarray}
which can be integrated for any invertible $T$ (we verified analytically that $T$ is indeed invertible). 
The second purpose is to impose the initial conditions. Fortunately (as we explicitly show below) the matrix $T \simeq 1$ with extremely good accuracy in the early time / sub-horizon regime. We can therefore simply disregard the departure of $T$ from the identity matrix in the early time regime, and perform  all the steps outlined in Subsection \ref{subsec:PS}. The initial conditions obtained in this way are extremely accurate. Finally,  the expressions (\ref{cal-O}) - (\ref{PS-OO}) for the observables  do not require that    $T = 1$.  Therefore, we do not need to explicitly  perform the transformation that sets $T$ to unity.

The procedure to obtain the matrices $T,K,\Omega^2$ in (\ref{S24-formal})  is a straightforward algebraic procedure, that we have outlined in Subsection  \ref{subsec:formal-deco}. The action (\ref{s2sca-explicit}) is of the form  (\ref{S2-formal}), and by comparing these two expressions we can immediately read off the explicit expressions of the matrices   $A, \dots , F$. We therefore also have the explicit expression for the action in terms of the dynamical modes, see eq.  (\ref{S22-formal}).   It is conceptually straightforward to perform the transformation (\ref{XtoD-explicit}) in this action, and obtain the explicit expressions for  $T,K,\Omega^2$. This leads to an expression for the matrix elements that is the scalar-sector counterpart of the expressions (\ref{KL})   and (\ref{omegaL}) that we had obtained for the tensor sector. However, these expressions (that we obtained using Mathematica) are extremely much longer 
and more involved than those in the tensor sector, and their full form is not particularly illuminating. For this reason  we do not report them here. All entries in these matrices are formally of the type~\footnote{The formal expression 
 (\ref{TKOM-formal}) has been written in a way that applies to all matrix elements, but not all the sums appearing in (\ref{TKOM-formal}) are nontrivial for all the matrix elements.  For instance, there is no square root in $\Omega_{11}^2$. Hence,   $\Omega_{11}^2$ is also formally of the the type (\ref{TKOM-formal}), but the sums inside the square root only contain the monomial $1$.  See  Appendix \ref{app-scalaraction} for the explicit forms of the various entries.}

\begin{equation}
\frac{\sum_i c_i p^{\alpha_i} }{\sum_j d_j p^{\alpha_j} }
\times \sqrt{ \frac{\sum_m {\tilde c}_m p^{\alpha_m} }{\sum_n {\tilde d}_n p^{\alpha_n } } } \;\;,
\label{TKOM-formal}
\end{equation}
where the sums are finite sums, and where  the coefficients $c_i, d_j, {\tilde c}_m , {\tilde d}_n$ are slowly evolving functions of time.

We expanded each of these coefficients in slow roll, performing the same two steps (\ref{subs-before-slowroll})   
and (\ref{subs-slowroll}) that we performed in the tensor sector. This leads to the expressions  (\ref{slowT}),  (\ref{slowK}),  (\ref{slowOm}) that we report in   Appendix \ref{app-scalaraction}, and that are the scalar-sector counterpart of eqs. (\ref{tensor-slow}).  We stress that the expressions given in the Appendix retain all the coefficients of (\ref{TKOM-formal}), and approximate each of them in slow roll. Therefore, these expressions are  very accurate at all times (namely for all values  of $p/H$).  These are the expressions that we used to integrate the equations  (\ref{eom-formal}) numerically. 
 
 We now discuss the asymptotic limits of the matrices $T,K,\Omega^2$ in the early time / sub-horizon and late-time / super-horizon regime. At early times, these evaluate to
 \begin{eqnarray}
p \gg H : 
 && T_{11} \simeq 1 + \frac{3 \epsilon^2 \gamma}{\left( 1 + \gamma \right)^2} \frac{H^2}{p^2} \;\;,\;\;
 T_{12}   \simeq - \frac{\sqrt{3 \gamma} \epsilon}{1+\gamma} \frac{H}{p} \;\;,\;\; 
  T_{22} \simeq 1 + \frac{3 \epsilon \gamma}{ 1 + \gamma } \frac{H^2}{p^2} \;\;, \nonumber\\
&& \frac{K_{12}}{a} \simeq - \frac{p}{\sqrt{3 \gamma}} \;\;, \nonumber\\
 && \frac{\Omega_{11}^2}{a^2} \simeq \frac{p^2}{3} \;\;,\;\; 
  \frac{\Omega_{12}^2}{a^2} \simeq \frac{2 \left( 1  + \gamma \right)}{\sqrt{3 \gamma}} p H \;\;,\;\; 
  \frac{\Omega_{22}^2}{a^2} \simeq \left( 1 - \frac{2}{\gamma} \right) p^2 \;\;,
\label{earlyTKOm}
  \end{eqnarray}
  where we recall that $T$ and $\Omega^2$ are symmetric, while $K$ is anti-symmetric.  These expressions have two properties worth noting. Firstly,  the asymptotic form of $\Omega^2$ shows that the scalar perturbations have a strong instability in the sub-horizon regime for $\gamma < 2$. We discuss this more in details in  Subsections \ref{subsec:initial} and  \ref{subsec:scalarsol}. Secondly, we see that   the matrix $T$  indeed  coincides with the identity in this regime, up to terms suppressed both by $H/p$ and by slow roll. The initial conditions are given by setting the modes in the initial adiabatic vacuum. This requires for instance that $\dot{\Omega} \ll \Omega^2$. We see from (\ref{earlyTKOm}) that $\frac{\dot{\Omega} }{\Omega^2 } = {\rm O } \left( \frac{H}{p} \right)$, while $T- \mathbb{I} = 
    {\rm O } \left( \epsilon \,  \frac{H}{p} \right)$. Therefore it is completely safe to approximate $T$ with the identity at the initial time.
    
However,  at horizon crossing and after it, $T$ departs from the identity. More in general, at late times we have 
 \begin{eqnarray}
 p \ll H : 
 && T_{11} \simeq 1 + \frac{2}{\gamma} \;\;,\;\;
 T_{12} \simeq - \frac{2 \sqrt{1+\gamma}}{\gamma \sqrt{\epsilon}} \;\;,\;\;
 T_{22} \simeq 1 + 2 \frac{1+\gamma}{\gamma \epsilon} \;\;, \nonumber\\
&& \frac{K_{12}}{a} \simeq \frac{2 \epsilon^{3/2} H}{3 \sqrt{1+\gamma}} \;\;, \nonumber\\
&& \frac{\Omega_{11}^2}{a^2} \simeq \frac{2 \left( - 2 + \gamma + \gamma^2 \right)}{\gamma} H^2 \;\;,\;\;
  \frac{\Omega_{12}^2}{a^2} \simeq \frac{4 \sqrt{1+\gamma}}{\gamma \sqrt{\epsilon}} H^2 \;\;,\;\;
  \frac{\Omega_{22}^2}{a^2} \simeq - \frac{4 \left( 1 + \gamma \right)}{\gamma \epsilon} H^2 \;\;.    
\label{lateTKOm}
   \end{eqnarray} 
  The asymptotic limits (\ref{earlyTKOm}) and (\ref{lateTKOm})  are significantly simpler than the  expressions (\ref{slowT}),  (\ref{slowK}),  (\ref{slowOm}).    However, they are not accurate at horizon crossing, and for this reason the expressions of Appendix \ref{app-scalaraction}  (which are valid at all times) are used  in the numerical integration of (\ref{eom-formal}).                                        
 
We conclude this Subsection with a note on the effect of the metric perturbations on the scalar system. To study their impact, we (i) artificially removed them from  (\ref{s2sca-explicit}), (ii)  integrated out the  only remaining non-dynamical mode $Y$, and (iii)  performed the transformation  (\ref{XtoD-explicit}). The resulting action   is  of the form   (\ref{S24-formal}) with $T= \mathbb{I} $ (this is true exactly, and not just in the slow roll or early/late time approximation). Therefore the transformation  (\ref{XtoD-explicit}) is the one that sets the kinetic matrix to unity in absence of metric perturbations.
As it is always the case in slow roll inflation, the effect of  metric perturbations is negligible in the sub-horizon regime, and indeed we could verify that the action obtained by the steps (i)-(ii)-(iii) coincides with the asymptotic early times  expressions (\ref{earlyTKOm})  to leading order in slow roll.

\subsection{Initial adiabatic solution}

\label{subsec:initial} 

In this Subsection we first of all   obtain an analytic approximate but very accurate solution for the  equation (\ref{eom-formal}) in the early time regime. This solution is characterized by some integration constants that we fix by imposing that the solution is initially in the positive frequency adiabatic vacuum, with the conditions  (\ref{wronskian}) and (\ref{wronskian2})       respected. This provides the initial conditions for the perturbations, that we use in Subsection   
\ref{subsec:scalarsol} when we numerically integrate the equation  (\ref{eom-formal}) at all times.

From the asymptotic early time expressions (\ref{earlyTKOm}) we obtain
\begin{eqnarray}
 \frac{k}{a} \gg H : 
 && \alpha = 
   \left( \begin{array}{cc}  
0 & - \frac{2 k}{\sqrt{3 \gamma}} \\
 \frac{2 k}{\sqrt{3 \gamma}} & 0 
\end{array} \right) + {\rm O } \left( a H \right)  \;\;,\;\;
 \beta = \left( \begin{array}{cc}  
 \frac{k^2}{3} & 0 \\
0 &  \frac{\gamma-2}{\gamma} k^2 \end{array} \right) + {\rm O } \left( a H k  \right)   \;\;,
\label{alpha-beta-early}
  \end{eqnarray}
for the two matrices that characterize the evolution equation  (\ref{eom-formal}). 

If we keep only the terms explicitly given in (\ref{alpha-beta-early}), we can guess that eq. (\ref{eom-formal}) should  
admit a solution characterized by $ \frac{\cal D'}{\cal D} = {\rm O } \left( k \right) \;,  \frac{\cal D''}{\cal D} = {\rm O } \left( k^2 \right) $. Indeed, such a solution is 
\begin{eqnarray}
{\cal D}^{\rm early}_{1j} & = & c_{1j} {\rm e}^{-i k \tau} +  c_{2j} {\rm e}^{i k \tau} + c_{3j} {\rm e}^{-i \frac{\sqrt{\gamma-2} k \tau}{\sqrt{3 \gamma}}}  + c_{4j} {\rm e}^{i \frac{\sqrt{\gamma-2} k \tau}{\sqrt{3 \gamma}}}  \;\;, \nonumber\\
{\cal D}^{\rm early}_{2j} & = & - \frac{i \sqrt{\gamma}}{\sqrt{3}} \left[  c_{1j} {\rm e}^{-i k \tau} -  c_{2j} {\rm e}^{i k \tau} \right]
+ \frac{i}{\sqrt{\gamma-2}} \left[   c_{3j} {\rm e}^{-i \frac{\sqrt{\gamma-2} k \tau}{\sqrt{3 \gamma}}}  - c_{4j} {\rm e}^{i \frac{\sqrt{\gamma-2} k \tau}{\sqrt{3 \gamma}}} \right] \;\;, \;\;  j=1,2 \;\;, 
\label{early-sol-1}
\end{eqnarray}
where $c_{ij}$ are integration constants. It is straightforward to verify that (\ref{early-sol-1}) indeed satisfies 
(\ref{eom-formal}) if only the terms explicitly given in (\ref{alpha-beta-early}) are retained. It is also straightforward to verify that, once the early time solution is inserted in (\ref{eom-formal}), the subdominant terms in (\ref{alpha-beta-early}) 
contribute to (\ref{eom-formal}) with terms that are suppressed by at least ${\rm O } \left( \frac{a H}{k} \right) \ll 1$ factors 
with respect to the contributions that we include.  This confirms that (\ref{early-sol-1}) solves the evolution equations in the early time / sub-horizon regime. Moreover, we can see that the system (\ref{eom-formal}) should admit $8$ complex integration constants. They are precisely given by the coefficients  $c_{ij}$ in (\ref{early-sol-1}). Therefore, the expression (\ref{early-sol-1}) is the most general early time  solution of eq.    (\ref{eom-formal}).  

As we anticipated in Subsection \ref{subsec:S2scal-explicit}, one linear combination of the modes   $\propto c_{3j}, c_{4j}$  is exponentially growing  in the solution (\ref{early-sol-1}) for $\gamma < 2$, signaling a tachyonic instability.     In general, a tachyonic instability is not an intrinsic pathology of a model, but ``simply'' an instability of a given background solution. If the instability is mild enough, it does not lead to any problem (as it is the case for the tachyonic instability in the left-handed tensor sector at sufficiently small $\gamma_{\rm in}$, see Section \ref{sec:tensor}). However, the present instability at $\gamma < 2$  is extremely strong. It takes place all throughout the sub-horizon regime, with a rate $\propto \frac{k}{a H}$, which is arbitrarily large the deeper inside the horizon a mode is. At any given moment, all modes in the deep sub-horizon regime are highly unstable.  
 
  In principle, the instability would be absent if we could set to zero  the coefficient of the linear combination    
that is exponentially growing in  (\ref{early-sol-1}) at $\gamma < 2$. Doing so, however,   we verified by direct inspection  that      the first Wronskian condition (\ref{wronskian}) cannot be satisfied for any $\gamma < 2$ (the different matrix components in that relation give incompatible requirements on the integration constants).  Therefore, already from this  condition   alone, we learn that it is not possible to avoid the instability  for $\gamma < 2$. Condition   (\ref{wronskian}) is obtained from the initial quantization. One may be tempted to simply postulate that quantization is impossible in that regime, and so set that coefficient to zero by hand. This would  however  be  a completely unjustified and acausal choice, as there is no causal reason why all sub-horizon modes should have at some given initial moment $t_{\rm in}$ a classical initial condition that avoids the tachyonic growth taking place  at $t > t_{\rm in}$. It would be equivalent to stating that the model $V= \frac{1}{2} m^2 \varphi^2$, with a large and negative $m^2$, is stable because, classically, the choice $\delta \varphi = 0$ can be made.  Thus, we conclude that the background solution obtained in Section \ref{sec:model} is invalid for  $\gamma < 2$. 

  We note that an instability inside the horizon takes place for some choice of parameters takes place  also in the related model of Chromo-Natural inflation  in \cite{Dimastrogiovanni:2012ew}. In that case the instability occurs  for a finite range  of $\frac{k}{a H}$ inside the horizon. We see that the instability in Gauge-flation is even stronger. However, both instabilities have the effect of invalidating the range of parameters in which they take place.

Let us now study  the stable $\gamma > 2$ region. As standard in inflation, we impose that the negative frequency terms in (\ref{early-sol-1}) are initially vanishing, $c_{2j} = c_{4j} = 0$. This corresponds to no quanta present at the initial time.~\footnote{This is well known for the single field case.  See \cite{Nilles:2001fg} for the multiple field case.} After imposing this, from a direct inspection we obtain that the 
conditions   (\ref{wronskian}) and  (\ref{wronskian2}) are satisfied \footnote{More precisely, we impose these conditions at $\tau = 0$; as we discussed in  Subsection \ref{subsec:PS}, if all the conditions   (\ref{wronskian}) and  (\ref{wronskian2}) are satisfied at some time, then they are satisfied at all times.}  if and only if    
\begin{eqnarray}
\vert c_{11} \vert^2 + \vert c_{12} \vert^2  =  \frac{3}{2 k \left( 1 + \gamma \right)} \;\;,\;\; 
\vert c_{31} \vert^2 + \vert c_{32} \vert^2  =  \frac{\sqrt{3 \gamma} \sqrt{\gamma-2}}{2 k \left( 1 + \gamma \right)} \;\;,\;\;
 c_{11} c_{31}^* + c_{12} c_{32}^* = 0 \;\;.
\label{sol-combo}
\end{eqnarray}

These relations do not uniquely specify   the remaining coefficients $c_{1i}$ and $c_{3i}$. This however was expected: as we discussed in Subsection \ref{subsec:PS}, if the matrix ${\cal D}$ satisfies   the Wronskian conditions, then the product ${\cal D} \times U$, where $U$ is unitary, also satisfies them. Under a unitary transformation, the coefficients 
$c_{1i}$ and $c_{3i}$  transform as 
\begin{equation}
c_{1j} \rightarrow c_{1k} \, U_{kj} \;\;,\;\; c_{3j} \rightarrow c_{3k} \, U_{kj} \;\;,
\label{U2}
\end{equation}     
and it is immediate to observe that the combinations in (\ref{sol-combo}) are left invariant by these transformations.
As we discussed in Subsection  \ref{subsec:PS}, the matrix $U$ is arbitrary, but unphysical, and it generalizes to the $N$ fields case the well known phase arbitrarity of the wave function of the  single field quantization.  There are $8$ real parameters in the $c_{1j},c_{3j}$ coefficients, and eqs. (\ref{sol-combo}) are $4$ real constraints. The space of allowed solutions is therefore a $4-$dimensional real space, which is the same dimension of the U(2) arbitrarity in (\ref{U2}). Therefore, we simply need to provide one solution of (\ref{sol-combo}), with the understanding that the other solutions will be related to it  by the arbitrary and unphysical transformation (\ref{U2}). An immediate solution of (\ref{sol-combo}) is provided by
\begin{equation}
c_{11} = \sqrt{\frac{3}{1+\gamma}} \, \frac{1}{\sqrt{2 k}} \;\;,\;\; c_{12} = 0 \;\;,\;\; 
c_{31} = 0 \;\;,\;\; c_{32} = \frac{\left( 3 \gamma \right)^{1/4} \left( \gamma - 2 \right)^{1/4}}{\sqrt{1+\gamma}} \,  \frac{1}{\sqrt{2 k}} \;\;.
\end{equation}

Inserting this solution in (\ref{early-sol-1}), and setting $\tau = 0$ as our initial time (changing the value of the initial time  
 in  (\ref{early-sol-1})  also corresponds to an unphysical U(2) transformation of ${\cal D}$), we obtain the initial conditions
for ${\cal D}$ and ${\cal D}'$:
\begin{equation}
\sqrt{2 k} \, {\cal D}^{\rm in} =   
 \left( \begin{array}{cc}
 \sqrt{\frac{3}{1+\gamma}} &  \frac{\left( 3 \gamma \right)^{1/4} \left( \gamma - 2 \right)^{1/4}}{\sqrt{1+\gamma}} \\ 
 - \frac{i\sqrt{\gamma}}{\sqrt{1+\gamma}} & \frac{i\left( 3 \gamma \right)^{1/4} }{ \left( \gamma - 2 \right)^{1/4} \sqrt{1+\gamma}} 
 \end{array} \right) \;\;,\;\;
 \sqrt{2 k}  {\cal D}^{'{\rm in} } = - i \, k \left( \begin{array}{cc}
      \sqrt{\frac{3}{1+\gamma}} &  \frac{ \left( \gamma - 2 \right)^{3/4}}{  \left( 3 \gamma \right)^{1/4}   \sqrt{1+\gamma}} \\ 
 - \frac{i\sqrt{\gamma}}{\sqrt{1+\gamma}} & \frac{i \left( \gamma - 2 \right)^{1/4}}{ \left( 3 \gamma \right)^{1/4}   \sqrt{1+\gamma}} 
 \end{array} \right) \;\;.
\label{Din}
            \end{equation}

These quantities are  the initial conditions for the numerical integration of the evolution equation (\ref{eom-formal})
the we perform in the next Subsection.

\subsection{Solution at all times, and power of $\zeta$  } 

\label{subsec:scalarsol}

We solve for the variables $\sqrt{2 k} {\cal D}_{ij}$. Starting from the initial conditions (\ref{Din}), we numerically integrate the equations of motion following what is formally written in (\ref{eom-formal}), where  the explicit expressions for the matrix elements are given in Appendix \ref{app-scalaraction}. We recall that these expressions are accurate in slow roll at all times (namely for all values of the ratio $p/H$). We integrate these equations after writing them in physical time, and in rescaled dimensionless quantities (\ref{tilde-combo}). The equations are solely written in terms of these rescaled quantities and of the rescaled physical momenta (\ref{p-tilde}).

From the solutions, we compute  the curvature perturbation on uniform density hypersurfaces, $\zeta$, that, in the spatially  flat gauge that we are using, is given by
\begin{equation}
\zeta \equiv - \frac{H}{\dot{\rho}} \delta \rho \;\; . 
 \label{def-zeta}
\end{equation}

\begin{figure}
\centering
\includegraphics[width=0.45\textwidth]{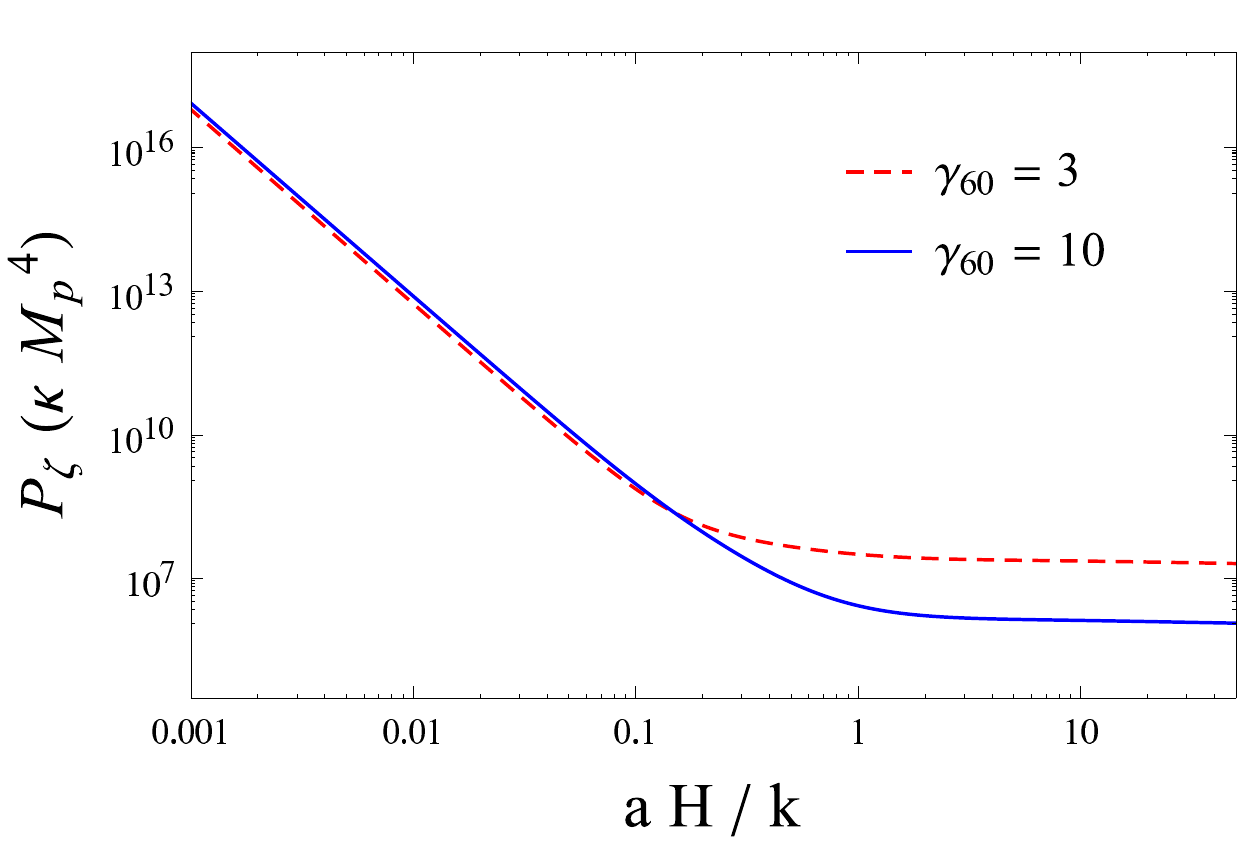}
\includegraphics[width=0.45\textwidth]{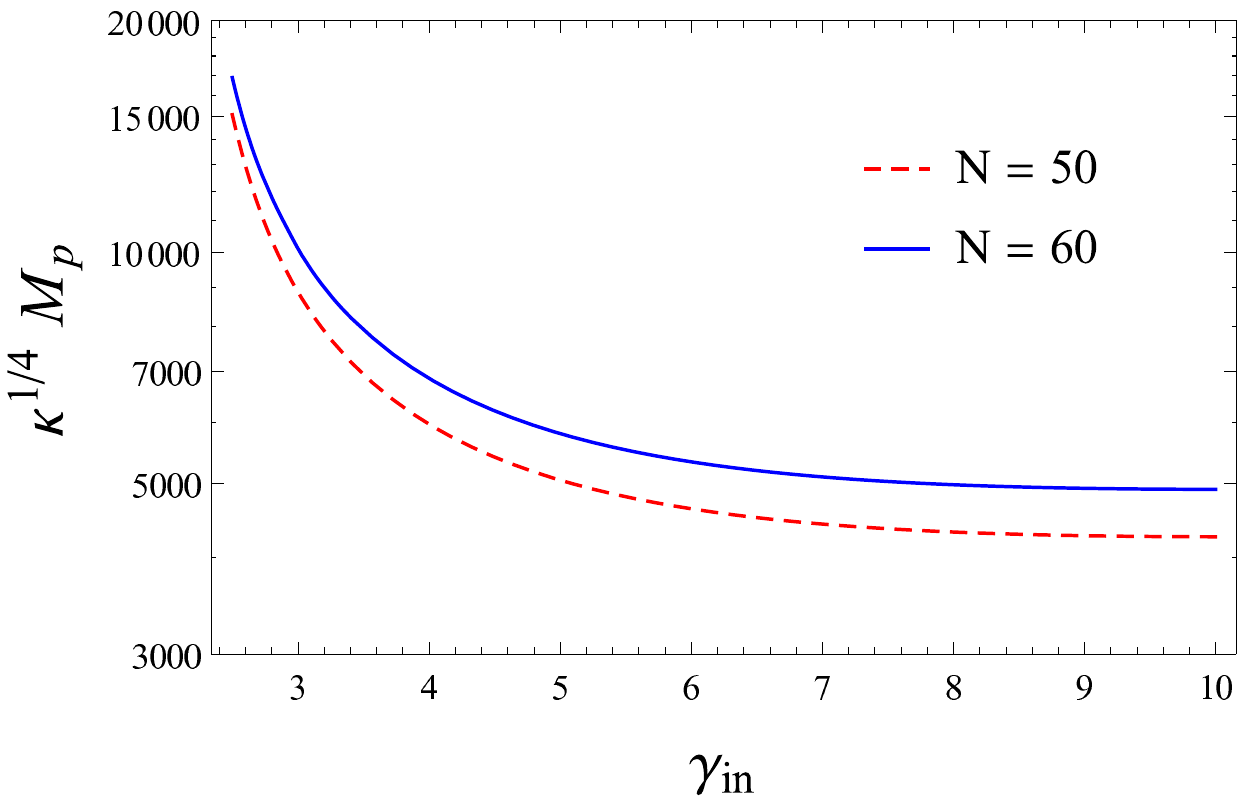}
\caption{ Left panel: time evolution of the power of $\zeta$ for a mode that leaves the horizon $60$ e-folds before the end of inflation, and for two different values of the variable $\gamma$ evaluated at that time (namely when $a H / k = 1$ in the Figure). Right panel: value  of the dimensionless quantity $\kappa^{1/4} M_p$ as a function of the initial value of $\gamma$, for $N=50$ and $N=60$ e-folds of inflation. This value is obtained by imposing the observed normalization 
 $P_\zeta =  2.2 \cdot 10^{-9}$  \cite{Ade:2013zuv} for the large scale modes leaving the horizon either $N=60$ or $N=50$ e-folds before the end of inflation. 
}     
\label{fig:Pzeta-kappa}
\end{figure}

Combining the expressions (\ref{Pzeta}) and (\ref{coeffzeta-super}), we obtain the super-horizon relation between the power of $\zeta$ and the     $\sqrt{2 k} {\cal D}_{ij}$ solutions:
\begin{eqnarray}
\!\!\!\!\!\!\!\! \!\!\!\!\!\!\!\!  P_\zeta \simeq \frac{1}{\kappa M_p^4} \, \frac{\left( 1 + \gamma \right)^2 {\tilde p}^2}{6 \pi^2 \gamma^2 \epsilon^4 } \sum_i \!\!\!\! & & \!\!\!\! \Bigg\vert - \frac{\sqrt{\epsilon}}{\sqrt{1+\gamma}} \left( 2 \sqrt{2 k } {\cal D}_{1i} + \frac{{\tilde \partial } \left[ \sqrt{2 k } {\cal D}_{1i} \right] }{ { \tilde \partial } a / a } \right) \nonumber\\ 
& & \quad \quad  \quad \quad  \quad \quad  \quad \quad  \quad \quad  + 2 \sqrt{2 k} {\cal D}_{2i} +   \frac{{\tilde \partial } \left[ \sqrt{2 k } {\cal D}_{2i} \right] }{ {\tilde \partial } a / a }   \Bigg\vert^2 \;\;,\;\; p \ll H \;\;. 
\label{PzetaD}
\end{eqnarray}
We verified numerically that this expression is accurate only $ \sim 10 $ e-folds after horizon crossing, due to the fact that the super horizon approximated results  (\ref{coeffzeta-super}) have been used.  In the left panel of  Figure \ref{fig:Pzeta-kappa} we show the time evolution of the power for a mode that leaves the horizon $60$ e-folds before the end of inflation, for two different choices of the parameter $\gamma$ evaluated at $60$ e-folds before the end of inflation.  In this Figure  and in the ones presented in the next Section,  the expressions  (\ref{coeffzeta-all}), rather than  (\ref{coeffzeta-super}), have been used. The expressions    (\ref{coeffzeta-all}) are valid at all times. The expressions  (\ref{coeffzeta-super}) are however significantly simpler, and accurate from  $ \sim 10 $ e-folds after horizon crossing onwards. 

In the left panel of Figure \ref{fig:Pzeta-kappa}, and in Figures \ref{fig:PLPR}, the power is given in units of $\frac{1}{\kappa M_p^4}$. The coefficient $\kappa$ can be finally obtained by imposing the power spectrum normalization $P_\zeta =  2.2 \cdot 10^{-9}$  \cite{Ade:2013zuv} for the large scale modes leaving the horizon either $N=60$ or $N=50$ e-folds before the end of inflation (according to the value specified in each plot shown). In the right panel of 
Figure  \ref{fig:Pzeta-kappa} we show the resulting value of $\kappa^{1/4} M_p$ (this is a dimensionless quantity)  as a function of $\gamma_{\rm in}$.

\section{Phenomenology}

\label{sec:pheno}

\begin{figure}
\centering
\includegraphics[width=0.45\textwidth]{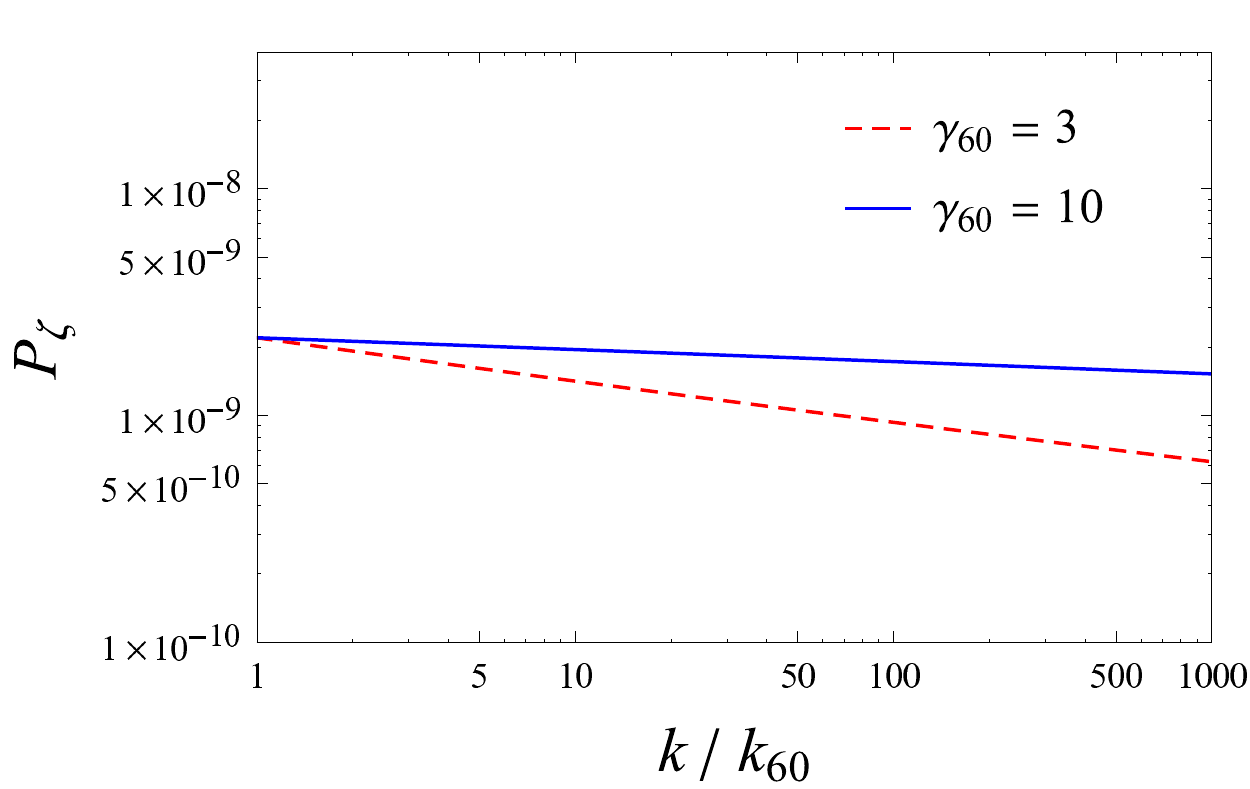}
\caption{ Power spectrum of $\zeta$ after freeze out. $k_{60}$ corresponds to the comoving  momentum of a mode leaving the horizon  $60$ e-folds before the end of inflation. We show the spectrum obtained for two different values of $\gamma$ at that moment.
}     
\label{fig:Pzeta} 
\end{figure}

Let us now discuss the phenomenological implications of the comparison between the results obtained in the previous two Sections for the tensor and scalar modes and the Planck results  \cite{Ade:2013zuv}. Figure \ref{fig:Pzeta} shows the power spectrum $P_\zeta$ for $N=60$ e-folds of inflation and   for two different values of $\gamma$ at that moment.

For each value of $k$ that has been used in the Figure, we show the value of  $P_\zeta$ obtained at $3$ e-folds after the horizon crossing; in each case we verified that by that time  the power had already saturated to its final freeze out value (this can also be seen in the example shown in Figures \ref{fig:PLPR} and \ref{fig:Pzeta-kappa}; we recall that $3$ efolds corresponds to $a H / k \simeq 20$). We  show the spectrum for a limited range of momenta, namely from $k=k_{60}$ (defined as the mode that leaves the horizon precisely $60$ e-folds of inflation) to $k = 10^3 k_{60}$.  However, given the mild (and smooth) scale dependence of $P_\zeta$ (typical of slow-roll inflation), the range of momenta shown is enough to determine the value of the spectral tilt $n_s$, defined~\footnote{To obtain $n_s$, we first compute $P_\zeta$ for a dense grid of values of $k$; we then interpolate $ \log P_\zeta$ as a function of $\log \frac{k}{k_{60}}$, and we differentiate this function. We verified that the grid is sufficiently dense, and that removing some of its points does not change the value of $n_s$.}  as $P_\zeta \propto k^{n_s-1}$. We see that the spectrum is more red at the smaller value of $\gamma_{\rm in}$ shown. 

\begin{figure}
\centering
\includegraphics[width=0.45\textwidth]{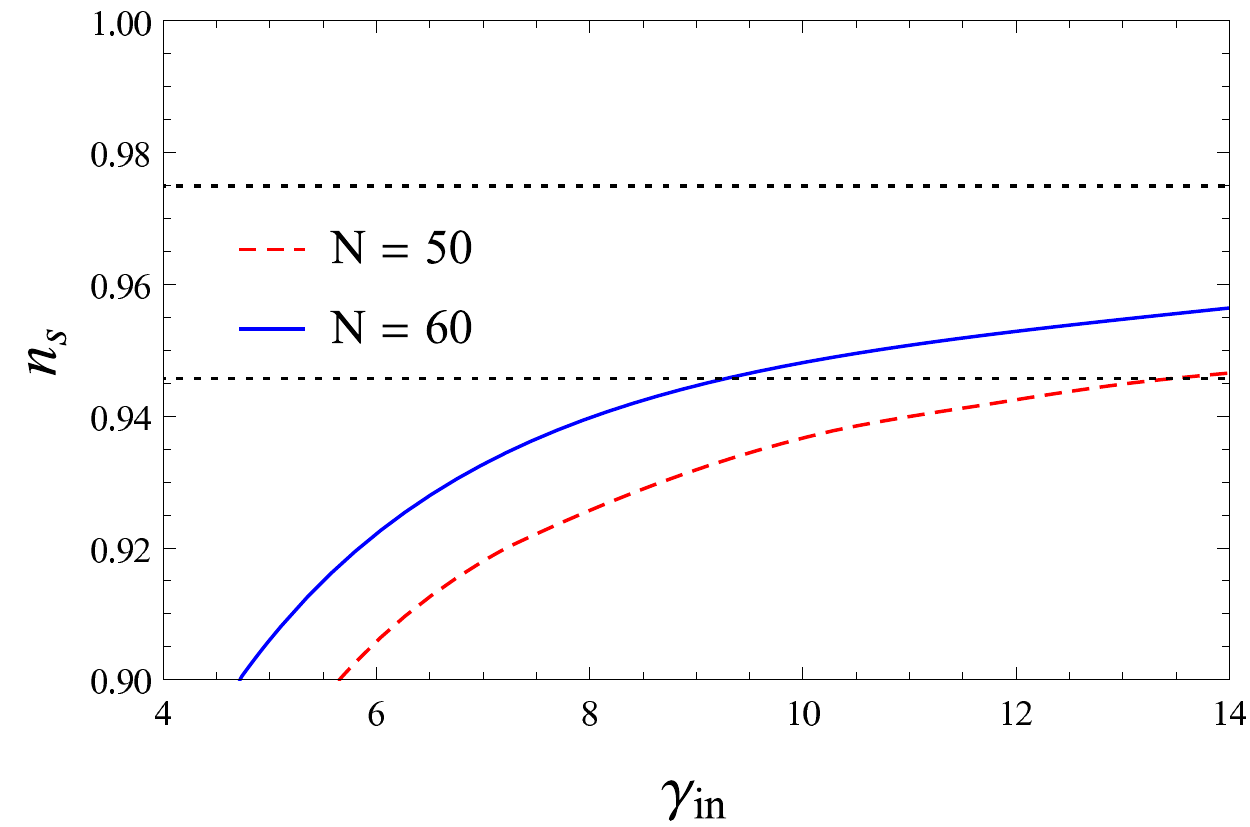}
\includegraphics[width=0.45\textwidth]{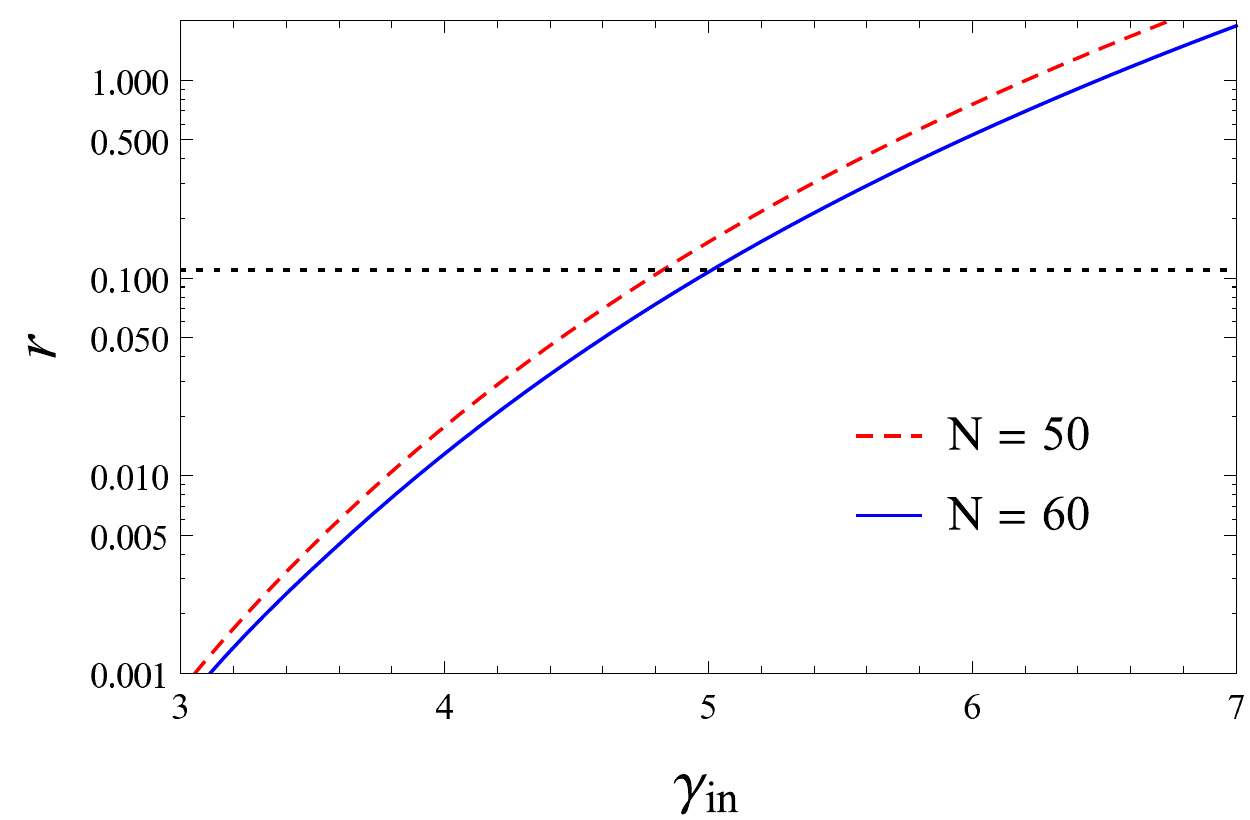}
\caption{ Spectral index $n_s$ and tensor-to-scalar ratio $r$ obtained in Gauge-flation for two different values of the number of e-folds $N$ and of the initial value of $\gamma$ (we recall that these two parameters completely characterize the model and the background evolution). Also shown are the  $95 \%$ CL Planck limits   \cite{Ade:2013zuv}    $0. 9457 < n_s < 0.9749$ and  $r < 0.11$. No value of these parameters is compatible with these limits, so that Gauge-flation is phenomenologically ruled out.   
}     
\label{fig:ns-r}
\end{figure}

Figure \ref{fig:ns-r}  is the main result of this work. The left panel shows the spectral index $n_s$  obtained for $N=50$ and $N=60$ e-folds of inflation, as a function of the initial value of $\gamma$ (evaluated at $50$ or $60$ e-folds before the end of inflation, respectively). Also the $95 \%$ CL Planck limits   \cite{Ade:2013zuv} are shown: $0. 9457 < n_s < 0.9749$.   The comparison excludes at  $95 \%$ CL all values $\gamma_{\rm in} \lta 13.5$ for $N=50$ and  all values $\gamma_{\rm in} \lta 9.3$ for $N=60$. ~\footnote{This confirms the observation made based on Figure \ref{fig:Pzeta} that the spectrum is redder at smaller $\gamma_{\rm in}$; we also see that, for fixed $\gamma_{\rm in}$, the spectrum is redder the smaller $N$ is (this is common in slow-roll inflation, since the departure from scale invariance is due to the slow roll parameters, that typically  increase during inflation).}  The  right panel of the Figure shows the value of the tensor-to-scalar ratio 
\begin{equation}
   r \equiv \frac{ P_L + P_R }{ P_\zeta } \;\; , 
\label{def-r}
\end{equation}
   as a function of the initial value of $\gamma$.  The   $95 \%$ CL Planck limit  \cite{Ade:2013zuv}  $r < 0.11$, rules out all values  $\gamma_{\rm in} \gta 4.8$ for $N=50$ and 
  $\gamma_{\rm in} \gta 5$ for $N=60$           ~\footnote{As we showed in Section \ref{sec:tensor}, the growth of $r$ with $\gamma_{\rm in}$ is due to the fact that the tachyonic instability in the left-handed tensor sector becomes stronger with increasing $\gamma$.}

We see that no value of the parameters leads to both  a sufficiently flat spectrum, and   sufficiently small tensor modes. We therefore conclude that Gauge-flation is ruled out by the CMB results. We employed the Planck limits given in the abstract of  \cite{Ade:2013zuv}, that correspond to a specific choice of priors. We can however see that the interval  of $\gamma_{\rm in}$ allowed by the $n_s$ bound  is rather far from the interval allowed by the $r$ bound, so that a small change of the allowed intervals (due to the different priors studied in  \cite{Ade:2013zuv}) does not impact our conclusion.


\section{Comparison with previous results}

\label{sec:comparison}

Refs. \cite{Maleknejad:2011jw} also performed a study of cosmological perturbations in 
Gauge-flation. To be concrete, all our remarks in this Section  refer to the most recent  version (namely, v$5$ posted on the archive) of the second paper in  \cite{Maleknejad:2011jw}. Ref.  \cite{Maleknejad:2011jw} reaches different conclusions from ours in the scalar sector. This crucially affects the phenomenological study of the model.  Their computation is in terms of gauge invariant variables, and it  is rather different from the procedure that we have outlined in Section \ref{sec:linpert}.  Therefore, a direct comparison between our and their procedure is not straightforward.  We refrain from commenting on their algebra. However, we can point out one crucial assumption and one crucial simplification done in \cite{Maleknejad:2011jw}, which we believe  invalidate their conclusions, and contribute to the disagreement between our and their results. 

A crucial assumption  of \cite{Maleknejad:2011jw},  which we believe to be incorrect is in the choice of initial conditions for $\gamma < 2 $. We have seen that $\gamma < 2$ leads to a strong tachyonic instability. Ref.  \cite{Maleknejad:2011jw} essentially removes the instability by setting to zero the coefficient of the exponentially large mode. Specifically, ref.  \cite{Maleknejad:2011jw} works in terms of the two modes ${\cal Q}$ and $\Phi$ (loosely speaking, they are the gauge invariant counterparts of our modes   $\delta Q$  and    $\phi$). They decompose these modes into ${\cal Q} = {\cal Q}_1+ {\cal Q}_2$ and  $\Phi = \Phi_1+ \Phi_2$, and each ${\cal Q}_i$ and $\Phi_i$ obeys a decoupled equation (their eqs. (V.68) and (V.69)) at asymptotically early times. The mode ${\cal Q}_1$ is related to $\Phi_1$ (and analogously for ${\cal Q}_2$ and $\Phi_2$) as explained in footnote $7$, so that this decomposition is in reality a rotation that diagonalizes the early times equations of motion. 
The eigenfrequencies coincide with those in our eq. (\ref{early-sol-1}), and the eigenfrequencies of the  modes   
${\cal Q}_2$ and $\Phi_2$ are tachyonic for $\gamma < 2$. 

Eq. (V.88) of   \cite{Maleknejad:2011jw} is the solution for the mode ${\cal Q}_2$, and eq.  (V.82) is the solution for $\Phi_2$. It is claimed in   \cite{Maleknejad:2011jw} that these are good solutions at all times, (we disagree on this, as we discuss below), but let us here only discuss the early time expansion of these solutions.  The first line of (V.88) is the solution for $\gamma <2$, while the second line is the solution  for $\gamma > 2$. Ref.   \cite{Maleknejad:2011jw} performs $2$ steps to obtain the initial conditions in these solutions.

Step 1: the coefficients $q_2$ and ${\tilde q}_2$  in the first line of  (V.88) are related to those in the second line, by stating that they are chosen  ``such that in both cases, ${\cal Q}_2$ has the same super horizon value.'' 

Step 2: The coefficient ${\tilde q}_2$ is set to zero based on the fact that, in the second line , this is the coefficient of the  negative frequency mode. 

Due to the identification in step 1, the choice done in step $2$ also sets  ${\tilde q}_2$ to zero in the first line  of (V.88). This coefficients multiplies the exponentially large mode at early times, and so this choice effectively removes the instability.  

Step 2 is correct. For $\gamma > 2$, there is no tachyonic instability, and the standard quantization indeed sets to zero the negative frequency modes (we are speaking of negative frequency, not negative squared frequency).  This is analogous to setting $c_{2j} = c_{4j}= 0$ in our eq. (\ref{early-sol-1}), and it is a standard choice of the vacuum in the sub-horizon regime. On the other hand, step 1 cannot be accepted. First of all, this step  is mathematically correct only for   ${\tilde q}_2 = 0$. If this is not the case, the first and second line of (V.88) do not coincide in the super-horizon regime  (this is not clarified in   \cite{Maleknejad:2011jw}). Moreover step 1 assumes that the solution (V.88) is valid at all times (as we discuss below, we disagree on this). These are however mathematical issues which are not the core of the problem.  The real problem with step 1 is that there is no physical reason why the initial conditions (in the sub-horizon regime) for one choice of the parameters should be given in such a way that they lead to a solution that, in the super-horizon regime, coincides with the solution obtained for another choice of parameters. Namely, according to  \cite{Maleknejad:2011jw},  if the background solution of Gauge-flation has $\gamma_{\rm in} = 1$, then the sub-horizon value for the scalar perturbations should be given such that the solution, once evolved to much later times, should coincide with the scalar perturbations obtained for a background solution with $\gamma_{\rm in} = 3$. 
For $N=60$ e-folds of inflation, the choice $\gamma_{\rm in} = 1$ corresponds to roughly $Q_{\rm in} \simeq 0.076 M_p$ and $\sqrt{\kappa} g \simeq 3,200 / M_p^2$, while the choice  $\gamma_{\rm in} = 3$ corresponds to roughly $Q_{\rm in} \simeq 0.049 M_p$ and $\sqrt{\kappa} g \simeq 12,000 / M_p^2$. Namely, the two cases are characterized by  completely different values of the parameters  and of the background  initial conditions, and 
there is no physical motivation for taking information from one case to set the initial conditions for the perturbations   in another case. It is  as unmotivated as for example having the model $V=\frac{1}{2} m^2 \varphi^2$, and setting the initial perturbations for $\delta \varphi$  in  the case $\varphi_{\rm in} = 1, m^2 = -2$, based on the fact that they should agree at late times with the perturbations for  $\varphi_{\rm in} = 3, m^2 = +4$. 

This is the reason why ref.  \cite{Maleknejad:2011jw} concludes that the $\gamma < 2$ region is stable.  We have shown that  this choice is inconsistent with a proper quantization, and physically unmotivated.

\bigskip

Also in the stable $\gamma >2$ region our results in the scalar sector disagree with those of  \cite{Maleknejad:2011jw}. 
Ref.   \cite{Maleknejad:2011jw} extends the decomposition  ${\cal Q} = {\cal Q}_1 + {\cal Q}_2$ and $\Phi = \Phi_1 + \Phi_2$ also to the late time regime. This decomposition must be  supplemented by two relations between these variables, not to artificially increase the number of degrees of freedom. These relations are given by footnote $7$ for the early time regime only. However, ref.   \cite{Maleknejad:2011jw} continues to use these modes for all the evolution, without providing a clear and explicit definition of how the ${\cal Q}_i$ and $\Phi_i$ modes are defined after the asymptotic early time regime. Maybe, one could argue that the solutions themselves encode how the modes are related to each other, and, although a clear definition of these quantities (prior the solutions are given) would be useful, perhaps this is just an issue of presentation of the results. Let us therefore disregard this issue, and rather let us   discuss more in details  the results given in  \cite{Maleknejad:2011jw} for  $\gamma > 2$. These  are given by  their eqs.   (V.81)-(V.82)-(V.87)-(V.88), and it is claimed in      \cite{Maleknejad:2011jw} that these results are the  solutions for ${\cal Q}_i$ and $\Phi_i$ at all times. The problem with  this statement is that  the equations solved by (V.81)-(V.82)-(V.87)-(V.88) are not the actual linearized equations for the perturbations of Gauge-flation. Specifically (V.81)-(V.82)-(V.87)-(V.88) are the solutions of   equations  (V.77)-(V.78)-(V.83)-(V.84). This set of equations is constructed so to coincide at early times with the early time equations derived in   \cite{Maleknejad:2011jw}  - these are equations   (V.68) and (V.69) - and so  to coincide at late times  with the  late time equations derived in   \cite{Maleknejad:2011jw}  - these are equations   (V.75) and (V.76).       Therefore, already in the derivation of ref.  \cite{Maleknejad:2011jw}, eqs.  (V.77)-(V.78)-(V.83)-(V.84) are not presented as the real equations for  the perturbations, but  as  equations that coincide with the real ones only in the deep sub-horizon and super-horizon regimes.            In  eqs.  (V.77)-(V.78)-(V.83)-(V.84) the variables  ${\cal Q}_i$ and $\Phi_i$ are fully decoupled. While - according to the derivation of  \cite{Maleknejad:2011jw} - these variables are decoupled in the deep sub- and super-horizon regimes, this is certainly not the case at horizon crossing. Therefore, assuming that eqs. 
 (V.77)-(V.78)-(V.83)-(V.84) are valid at all times trivializes all the dynamics, particularly in the most crucial times when the mixing may be expected to be mostly relevant (or, at the very least, where no $p \gg H$ nor $p \ll H$ approximation can be made). Ref.  \cite{Maleknejad:2011jw} does not study the accuracy of (V.77)-(V.78)-(V.83)-(V.84),  neither by comparing their solutions with the ones that can be obtained by numerically solving the real system of equations, nor by performing an analytic comparison of this set of equations with the real ones. 
 
  We stress that all entries in the equations that we have solved numerically coincide (at leading order in slow roll) with the exact ones at all times (for all values of $p/H$).   Our results disagree with those of   \cite{Maleknejad:2011jw}. This, together with the lack of a proper justification of eqs.  (V.77)-(V.78)-(V.83)-(V.84),  leads us to the conclusion that this set of equations does not correctly characterize the dynamics of the scalar perturbations of the model.



\section{Conclusions}

 \label{sec:conclusions}

Gauge-flation has the interesting peculiarity that it is a model of inflation driven by a vector field without ghost instabilities.
It has been shown that the model is related to a specific limit of Chromo-natural inflation  \cite{SheikhJabbari:2012qf}. This identification has been shown to be accurate at the background level  \cite{Adshead:2012qe,SheikhJabbari:2012qf}. However, as we discussed in the Introduction, it may be possible that integrating out the axion of 
 Chromo-natural inflation  leads to appreciable  differences between the perturbations in the two models. This possibility appears to be corroborated by the existing literature. The perturbations of Chromo-natural inflation have been studied by a number of recent works \cite{Dimastrogiovanni:2012st,Dimastrogiovanni:2012ew,Adshead:2013qp,Adshead:2013nka}. Their results agree with each other when they overlap:  the model is stable only if the vector field is sufficiently heavy, which is encoded by the condition  (\ref{cn-instab}). In the stable region, there is no value of parameters leading to sufficiently flat scalar spectrum and sufficiently small tensor modes. Perturbations in Gauge-flation were studied in \cite{Maleknejad:2011jw}, which did not report any unstable region, and which obtained a spectrum of scalar perturbations significantly bluer than those in Chromo-natural inflation.
 
Our results for the perturbations in Gauge-flation disagree with those of  \cite{Maleknejad:2011jw}, and confirm that the identification \cite{SheikhJabbari:2012qf} between the two models is accurate also at the perturbative level. Firstly, we have seen that the scalar perturbations of Gauge-flation are highly unstable when $\gamma \equiv \frac{g^2 Q^2}{H^2} < 2$. This condition  coincides with the result obtained for Chromo-natural inflation \cite{Dimastrogiovanni:2012ew}, see eq. (\ref{cn-instab}). The instability was discussed in  \cite{Dimastrogiovanni:2012ew} for the 
 Chromo-natural formulation of the model: it is first of all not an instability of the model, but ``only'' of the inflationary background solution. A background instability typically manifests itself through the presence of tachyonic modes, and this is also the case for these two models. The instability can be most easily explained in the Chromo-natural formulation  \cite{Dimastrogiovanni:2012ew}:  for any given value of $H$, the axion potential is too steep to drive inflation by itself.   Chromo-natural inflation realizes a slow-roll axion  evolution by coupling it with a gauge field. The mechanism requires that the gauge field is non-Abelian ($g\neq 0$) and it has a vev ($Q \neq 0$), and so - a posteriori - it is not surprising that, for any given $H$, the mechanism cannot work at arbitrarily small $g \, Q$. 

Secondly, let us compare the results for the perturbations in the two models in the stable $\gamma > 2$ region. For Gauge-flation, as shown in Fig.~6, both the tilt of the scalar power spectrum ($n_{s}$) and the tensor-to-scalar ratio ($r$) are growing functions of $\gamma_{\rm in}$. We find that the $95\%$ CL Planck limits \cite{Ade:2013uln} on the spectral index ($0.9457 < n_{s} < 0.9749$) requires at $95\%$ CL   $\gamma_{\rm in} \geq 13.5$ for $N = 50$ and  $\gamma_{\rm in} \geq 9.3$ for $N = 60$. The $95\%$ CL Planck limit on the tensor-to-scalar ratio $r < 0.11$, requires instead  $\gamma_{\rm in}\leq 4.8$ for $N = 50$ and $\gamma_{\rm in}\leq 5$ for $N = 60$. For Chromo-natural inflation, we refer to the parameter scan performed in ref. \cite{Adshead:2013nka}. As shown in their Figure 12, the Planck bounds on the scalar  spectral tilt translate into the limits $6.5 \lta \gamma_{\rm in} \lta 16$ for a number of e-folds between $50$ and $60$. Ref.  \cite{Adshead:2013nka} does not provide a direct plot (or equation) of $r$ as a function of $\gamma_{\rm in}$. However, we can approximately deduce a bound by comparing their Figures 12 and 13. From Figure 13, we see that $r$ is compatible with the Planck bound only when $n_s \lta 0.935$. From Figure 12 we see that this limit on $n_s$ translates into $\gamma_{\rm in} \lta 5$. We recall that the statement of equivalence between the two models is that Gauge-flation should correspond to a specific limit of Chromo-natural inflation. Therefore there is a large fraction of parameters in Chromo-natural inflation that do not correspond to Gauge-flation. Figures $12$ and $13$ of   \cite{Adshead:2013nka} scan also over these parameters. So, if we could restrict the results in the two Figures to only those regions of parameters compatible with Gauge-flation, we may find more tighter bounds for $\gamma_{\rm in}$. Therefore our result $\gamma_{\rm in} \gta {\rm O } \left( 10 \right)$ from the spectral index is compatible with that obtained in Chromo-natural inflation, while the 
result obtained in  \cite{Maleknejad:2011jw} is significantly bluer (see their Figure 8). Also the limits from $r$ obtained 
here are in good agreement with that of Chromo-natural inflation. 
 
To conclude, we have applied the formalism that we have developed  in our study \cite{Dimastrogiovanni:2012ew} of Chromo-natural inflation to the model of Gauge-flation. Gauge-flation was found to be equivalent to Chromo-natural inflation plus corrections \cite{SheikhJabbari:2012qf}, and the corrections were shown to be negligible at the background level  \cite{Adshead:2012qe,SheikhJabbari:2012qf}. The situation was less clear at the perturbative level. Our study has shown that the analogy persists also at this level. Both models are unstable at $\gamma  < 2$, and have too red a scalar spectrum at small $\gamma_{\rm in}$. In both cases, raising  $\gamma_{\rm in}  $ to a level that is compatible with the Planck bounds   \cite{Ade:2013uln} on $n_s$ results in too large a gravity wave signal. Therefore, identically to what happens for Chromo-natural inflation, also Gauge-flation is ruled out by the CMB data.

\vspace{.2 cm}

{\bf \large Acknowledgements}

\bigskip

\noindent  
We thank   A.~Maleknejad and M.~M.~Sheikh-Jabbari for correspondence on their work \cite{Maleknejad:2011jw} and on our Section \ref{sec:comparison}. This work is supported in part by DOE grant DE-FG02-94ER-40823 at the University of Minnesota. MP would like to thank the University of Padova, INFN, Sezione di Padova, and the Cosmology Group at the Department of Theoretical Physics of the University of Geneva for their friendly hospitality and for partial support during his sabbatical leave.

\appendix
\numberwithin{equation}{section}

\section{Scalar action in slow roll approximation}

\label{app-scalaraction}

Eq. (\ref{s2sca-explicit}) is the action of the scalar perturbations of Gauge-flation, in the gauge discussed in Subsection \ref{subsesec:gauge}. After integrating out the non-dynamical modes and  performing the transformation (\ref{XtoD-explicit}), we obtain an action of the form (\ref{S24-formal}), where each element in the matrices $T,K,\Omega^2$ is of the form (\ref{TKOM-formal}). 

We compute the slow roll approximation of each coefficient entering in  (\ref{TKOM-formal}), by performing the two steps outlined in eqs. (\ref{subs-before-slowroll})   and (\ref{subs-slowroll}).                                We obtain
\begin{eqnarray}
T_{11}  & \simeq & 1 + \frac{6 \epsilon^2 \gamma H^2}{2 \left( 1 + \gamma \right)^2 p^2 + 3 \epsilon^2 \gamma^2 H^2} \;\;, \nonumber\\
T_{12} = T_{21} & \simeq & - 2 \sqrt{3} \epsilon \frac{\sqrt{\gamma \left( 1 + \gamma \right)} \sqrt{ \left( 1 + \gamma \right) p^2 + 3 \epsilon \gamma H^2} H}{2 \left( 1 + \gamma \right)^2 p^2 + 3 \epsilon^2 \gamma^2 H^2} \;\;, \nonumber\\
T_{22} & \simeq & 1 +  \frac{6 \epsilon \gamma \left( 1 + \gamma \right)  H^2}{2 \left( 1 + \gamma \right)^2 p^2 + 3 \epsilon^2 \gamma^2 H^2} \;\;, 
\label{slowT}
\end{eqnarray}
for the first matrix, 
\begin{eqnarray}
K_{11} = K_{22} & = & 0 \;\;,  \nonumber\\
\frac{K_{12}}{a} = - \frac{K_{21}}{a} & \simeq & - \frac{2 \left( 1 + \gamma \right)^3 p^4 + 3 \epsilon \gamma \left( 1 + \gamma \right)^2 p^2 H^2 - 6 \epsilon^4 \gamma^3   H^4}{ \sqrt{3 \gamma \left( 1 + \gamma \right)} \sqrt{\left( 1 + \gamma \right) p^2 + 3 \epsilon \gamma H^2} \left[ 2 \left( 1 + \gamma \right)^2 p^2 + 3 \epsilon^2 \gamma^2 H^2 \right]} \;\;, 
\label{slowK}
\end{eqnarray}
for the second matrix, and
\begin{eqnarray}
\frac{\Omega_{11}^2}{a^2} & \simeq & 
\frac{p^2}{3} + 2 \left( 1 + \gamma \right) H^2 - \epsilon^2 \gamma H^2 \frac{4 \gamma \left( 1 + \gamma \right)^2 p^4-24 \left( 1 + \gamma \right)^2 p^2 H^2 + 36 \epsilon^2 \gamma^2 H^4}{\left[ 2 \left( 1 + \gamma \right)^2 p^2 + 3 \epsilon^2 \gamma^2 H^2 \right]^2} \;\;,
\nonumber\\
\frac{\Omega_{12}^2}{a^2} = \frac{\Omega_{21}^2}{a^2}  & \simeq & 
\frac{8\sqrt{1+\gamma}H}{\sqrt{3 \gamma} \sqrt{\left( 1 + \gamma \right) p^2 + 3 \epsilon \gamma H^2} \left\{ 4 \left( 1 + \gamma \right)^5 p^6 + 3 \epsilon \gamma \left[ 2 \left( 1 + \gamma \right)^2 p^2 + 3 \epsilon^2 \gamma^2 H^2 \right]^2 H^2 \right\}} \times \Bigg[  \left( 1 + \gamma \right)^6 p^8
\nonumber\\
& &          
\!\!\!\!\!\! \!\!\!\!\!\! 
 + 3 \epsilon \gamma \left( 1 + \gamma \right)^4 \left( 1 + 2 \gamma \right) p^6 H^2 + \frac{9}{4} \epsilon^2 \gamma^2 \left( 1 + \gamma \right)^3 \left( 4 \gamma - 5 \right) p^4 H^4 - 27 \epsilon^3 \gamma^3 \left( 1 + \gamma \right)^2 p^2 H^6 + \frac{81}{2} \epsilon^5 \gamma^5 H^8 \Bigg] \;\;,
\nonumber\\
\frac{\Omega_{22}^2}{a^2} \!\! \!\!  & \simeq & \!\! \!\!
 \frac{\gamma-2}{\gamma} p^2 + \epsilon H^2 \left( 1 + \gamma \right) \frac{ \left(    \gamma - 6  \right) \left( 1 + \gamma \right)^3 p^6 + 9 \gamma \left( 1 + \gamma \right)^3 p^4 H^2 + 18 \epsilon \gamma^2 \left( 1 + \gamma \right)^2 p^2 H^4 - 27 \epsilon^3 \gamma^4 H^6}{ \left( 1 + \gamma \right)^5 p^6 +   \frac{ 3 }{ 4 }   \epsilon \gamma \left[ 2 \left( 1 + \gamma \right)^2 p^2 + 3 \epsilon^2 \gamma^2 H^2 \right]^2 H^2} \;\;, \nonumber\\
\label{slowOm}
\end{eqnarray}
for the third matrix.

\section{Expression for $\zeta$ in terms of ${\cal D}_{ij}$}

\label{app-zetaD}

We are interested in the curvature perturbation on uniform density hypersurfaces $\zeta$, that, in the spatially  flat gauge that we are using is given by
\begin{equation}
\zeta \equiv - \frac{H}{\dot{\rho}} \delta \rho = \frac{\delta \rho}{6 \left[  \left( \dot{Q} + H Q \right)^2    + g^2 Q^4 \right] } 
\simeq \frac{\delta \rho}{6 H^2 Q^2 \left( 1 + \gamma \right) } \;\;,
\end{equation}
where the background equations have been used in the first equality, while the final expression is given at leading order in slow roll.

The variation of the energy is obtained by perturbing to linear order the definition $\rho \equiv - T^0_0$, and it reads
  \begin{eqnarray}
\delta \rho & = & 3 \frac{\left( a Q \right)'}{a^3} \left( 1 + \kappa g^2 Q^4 \right) \delta Q' - k^2 \frac{\left( a Q \right)'}{a^3} \left( 1 + \kappa g^2 Q^4 \right) M'
\nonumber\\ 
&& \!\!\!\!\!\!\!\!\!\!
+ 3 \left[ 2 \, g^2 Q^3 + \left( 1 + 3 \, \kappa g^2 Q^4 \right) \frac{a'^2}{a^4} Q + \left( 1 + 5 \, \kappa g^2 Q^4 \right) \frac{a' Q'}{a^3} +2 \, \kappa g^2 Q^3 \frac{Q'^2}{a^2} \right] \delta Q
\nonumber\\
&& \!\!\!\!\!\!\!\!\!\!
- k^2 \left[ 2 \, g^2 Q^3 + \left( 1 + 3 \, \kappa g^2 Q^4 \right) \frac{a'^2}{a^4} Q + \left( 1 + 5 \, \kappa g^2 Q^4 \right) \frac{a' Q'}{a^3} +2 \, \kappa g^2 Q^3 \frac{Q'^2}{a^2} \right] M
\nonumber\\
&& \!\!\!\!\!\!\!\!\!\!
+  k^2 \, \frac{\left( a Q \right)'}{a^3} \left( 1 + \kappa g^2 Q^4 \right) Y + 3 \, \frac{\left( a Q \right)'^2}{a^4} \left( 1 + \kappa g^2 Q^4 \right) \Phi \; .
\end{eqnarray}

We express the non-dynamical variables $Y$ and $\Phi$ in terms of the dynamical variables.~\footnote{Specifically, we impose eqs. (\ref{constraint-sol}), namely the solutions of the constraint equations. The explicit form of the matrices $D$, $E$, $F$ that appear in  (\ref{constraint-sol}) are obtained as explained in the paragraph before eq. (\ref{TKOM-formal}).} In this way, we express  $\delta \rho$  as a linear combination of $\delta Q$, $M$, and their first derivative. We expand to leading order in slow roll the coefficients of this expansion, and we obtain
\begin{eqnarray}
\delta \rho & \simeq & 
\frac{4 \left( 1 + \gamma \right)^{3/2} H M_p \sqrt{\epsilon}}{2 \left( 1 + \gamma \right)^2 p^2 + 3 \epsilon^2 \gamma^2 H^2} \Bigg\{ 3 H \left[ \left( 3 + \gamma \right) p^2 + 9 \gamma H^2 \right] \delta Q 
- a^2 p^2 H \left[ \left( 2 + \gamma \right) p^2 + 9 \gamma H^2 \right] M  \nonumber\\
&& \quad\quad\quad\quad  \quad\quad\quad\quad  \quad\quad\quad\quad 
+ \frac{3}{a} \left[ p^2 + 3 \gamma H^2 \right] \delta Q' - 3 a \gamma p^2 H^2 M' \Bigg\} \;\;. 
\end{eqnarray}

The variable $\zeta$ is our observable in the scalar sector, and we are  interested in its power spectrum. When comparing with (\ref{observable}), we have $X_1 =  \delta Q $ and $X_2 = M$, and therefore
\begin{equation}
\zeta = c_1 \, \delta Q + c_2 M + d_1 \delta Q' + d_2 M' \;\;,
\end{equation}
with
\begin{eqnarray}
& & c_1  \simeq   
\frac{2 \left( 1 + \gamma \right)^{3/2} }{ M_p \sqrt{\epsilon} } \; 
\frac{\left( 3 + \gamma \right) p^2 + 9 \gamma H^2}{2 \left( 1 + \gamma \right)^2 p^2 + 3 \epsilon^2 \gamma^2 H^2}   \;\;\; , \;\;\;  
c_2  \simeq   
- 2 a^2 p^2 \frac{ \left( 1 + \gamma \right)^{3/2} }{ 3 M_p \sqrt{\epsilon} } \; \frac{\left( 2 + \gamma \right) p^2 + 9 \gamma H^2}{2 \left( 1 + \gamma \right)^2 p^2 + 3 \epsilon^2 \gamma^2 H^2}      \;\; ,  \nonumber\\ 
& & 
d_1  \simeq   \frac{2 \left( 1 + \gamma \right)^{3/2}}{a H M_p \sqrt{\epsilon}} \; \frac{     p^2 + 3 \gamma H^2}{2 \left( 1 + \gamma \right)^2 p^2 + 3 \epsilon^2 \gamma^2 H^2}   \;\;\; ,  \;\;\; 
d_2  \simeq   - \frac{2 a H \gamma \left( 1 + \gamma \right)^{3/2}}{M_p \sqrt{\epsilon}} \; \frac{p^2}{ 2 \left( 1 + \gamma \right)^2 p^2 + 3 \epsilon^2 \gamma^2 H^2}        \;\; .   
\end{eqnarray}

Using eqs. (\ref{cal-O}) and (\ref{PS-OO}) we can see that the correlator of $\zeta$ is related to the matrix elements ${\cal D}$ by
\begin{equation}
P_{\zeta } = \frac{k^2}{4 \pi^2} \sum_{i} \left\vert \left( c_j {\cal M}_{jl} + d_j {\cal M}_{jl}' \right) \sqrt{2 k} {\cal D}_{li} + d_j {\cal M}_{jl} \sqrt{2 k} {\cal D}_{li}' \right\vert^2 \;\;. 
\label{Pzeta}
\end{equation}
which is the scalar-sector  analogous of the expression (\ref{PLR}).

We have
\begin{eqnarray}
& & \!\!\!\!\!\!\!\! 
 c_j {\cal M}_{jl} + d_j {\cal M}_{jl}'   =  
 \frac{\sqrt{2/3} \sqrt{1+\gamma}}{\left[ 2 \left( 1 + \gamma \right)^2 p^2 + 3 \gamma^2 \epsilon^2 H^2 \right] a M_p \sqrt{\epsilon} } 
 \nonumber\\  & & 
  \quad\quad\quad\quad  \times 
\left\{ -  \left[ \left( 2 + \gamma \right) \epsilon p^2 + 6 \epsilon \gamma H^2 \right] ,\,
\frac{\sqrt{1+\gamma}}{\sqrt{3 \gamma}} \frac{\left( 1 + \gamma \right) \left( 2 + \gamma \right) p^4 + 3 \gamma \left( 1 + \gamma \right) H^2 p^2 + 18 \gamma^2 \epsilon H^4}{H \sqrt{\left( 1 + \gamma \right) p^2 + 3 \gamma \epsilon H^2} } \right\}_l \;\;,  \nonumber\\ 
& & \!\!\!\!\!\!\!\! 
 d_j {\cal M}_{jl}       =  
 \frac{\sqrt{2/3} \sqrt{1+\gamma}}{\left[ 2 \left( 1 + \gamma \right)^2 p^2 + 3 \gamma^2 \epsilon^2 H^2 \right] H  a^2 M_p \sqrt{\epsilon} } 
\left\{ -  \left[ \left( 1 + \gamma \right)  p^2 + 3 \epsilon \gamma H^2 \right] ,\,
\sqrt{3 \gamma \left( 1 + \gamma \right) } H \sqrt{ \left( 1 + \gamma \right) p^2 + 3 \gamma \epsilon H^2 } \right\}_l \;\;. 
 \nonumber\\
\label{coeffzeta-all}
\end{eqnarray}

In the super-horizon regime, these expressions approximate to
\begin{eqnarray}
 c_j {\cal M}_{jl} + d_j {\cal M}_{jl}'  & \simeq  &  \frac{2 \sqrt{2} \left( 1+\gamma \right)}{\sqrt{3} \, \gamma} \frac{1}{ \epsilon^2 a M_p} \left\{ - \frac{\sqrt{\epsilon}}{\sqrt{1+\gamma} } , \, 1\right\}_l \;\; ,  \nonumber\\ 
 d_j {\cal M}_{jl}   & \simeq  &  \frac{1}{H a}  \frac{  \sqrt{2} \left( 1+\gamma \right)}{\sqrt{3} \, \gamma} \frac{1}{ \epsilon^2 a M_p } \left\{ - \frac{\sqrt{\epsilon}}{\sqrt{1+\gamma}} , \, 1 \right\}_l \;\; , \;\; p \ll H  \;\; . 
\label{coeffzeta-super}
 \end{eqnarray}
These are the expressions that we use in the main text in  eq. (\ref{PzetaD}). We verified numerically that these expressions provide the correct result for $\zeta$ starting from $\sim 10$ e-folds after horizon crossing. The freeze out of the power however occurs soon after horizon crossing, as can be observed by using the expressions (\ref{coeffzeta-all}), which are valid at all times (these are the expressions used in the left panel of  Figure \ref{fig:Pzeta-kappa}).

\end{document}